\documentclass[prb,aps,twocolumn,amsmath,amssymb]{revtex4}
\usepackage{graphicx,bm}
\usepackage{dcolumn}

\begin{document}

\title{A Renormalization-Group Study of Interacting Bose-Einstein condensates:\\
Absence of the Bogoliubov Mode below Four ($T>0$) and Three ($T=0$) Dimensions}

\author{Takafumi Kita}
\affiliation{Department of Physics, Hokkaido University, Sapporo 060-0810, Japan}

\begin{abstract}
We derive exact renormalization-group equations for the $n$-point vertices ($n=0,1,2,\cdots$) of interacting single-component Bose-Einstein condensates based on the vertex expansion of the effective action. 
They have a notable feature of automatically satisfying Goldstone's theorem (I),
which yields the Hugenholtz-Pines relation $\Sigma(0)-\mu=\Delta(0)$ as the lowest-order identity. 
Using them, it is found that the anomalous self-energy $\Delta(0)$ vanishes below $d_{\rm c}=4$ ($d_{\rm c}=3$) dimensions 
at finite temperatures (zero temperature),
contrary to the Bogoliubov theory predicting a finite ``sound-wave'' velocity $v_{\rm s}\propto [\Delta(0)]^{1/2}>0$.
It is also argued that the one-particle density matrix $\rho({\bm r})\equiv \langle \hat\psi^\dagger({\bm r}_1)\hat\psi({\bm r}_1+{\bm r})\rangle$  
for  $d<d_{\rm c}$ dimensions
approaches the off-diagonal-long-range-order value $N_{\bm 0}/V$ asymptotically as $r^{-d+2-\eta}$
with an exponent $\eta>0$.
The anomalous dimension $\eta$ at finite temperatures is predicted to behave for $d=4-\epsilon$ dimensions ($0<\epsilon\ll 1$)
as $\eta\propto\epsilon^2$.
Thus, the interacting Bose-Einstein condensates are subject to long-range fluctuations similar 
to those at the second-order transition point, and their excitations in the one-particle channel are distinct from 
the Nambu-Goldstone mode with a sound-wave dispersion in the two-particle channel.
\end{abstract}

\maketitle

\section{Introduction}

According to the Hugenholtz-Pines theorem\cite{HP59} and first proof of Goldstone's theorem\cite{GSW62,Weinberg96}
(refered to as {\it Goldstone's theorem (I)} hereafter),
Bose-Einstein condensates should embrace a branch of gapless (or massless) excitations in the one-particle channel.
It has long and widely been accepted\cite{Beliaev58,AGD63,GN64,HM65,FW72,SK74,WG74,Griffin93,Leggett01,PS03,Andersen04,OKSD05,WM13} that  the branch is identical with the mode predicted by the weak-coupling Bogoliubov theory,\cite{Bogoliubov47} which exhibits a linear dispersion relation at long wavelengths.
It will be shown here, however, that
the Bogoliubov theory is justified only above the critical dimensions $d_{\rm c}=4$ ($d_{\rm c}=3$) at finite temperatures (zero temperature),
below which the massless branch transforms into
critical fluctuations similar to those at the second-order transition point.
That this can be the case may be convinced by the following observation.

Second-order transitions are subject to long-range fluctuations in the order parameter characterized
by power-law behaviors with critical exponents in various physical quantities.\cite{WF72,Wilson72,Domb,Wilson74,Fisher74,SKMa,Amit,Justin96}
Mathematically, they are caused by vanishing of the mass term in the relevant Green's function $G$.
Specifically, the mean-field behavior $G^{(0)}(k)\propto k^{-2}$ ($k$: wavenumber) at the transition point
transforms into a distinct power-law behavior $G(k)\propto k^{-2+\eta}$ with
an exponent $\eta>0$ through the renormalization process of removing infrared divergences caused by the interaction.
For Bose-Einstein condensates, on the other hand, the absence of the mass term persists
down below the transition temperature. Indeed, non-interacting Green's function $G^{(0)}(k)$
behaves as $k^{-2}$ owing to vanishing of the chemical potential $\mu$ (i.e., the mass term).
Hence, it is natural to expect that switching on the interaction will result in the development of some exponent $\eta$ in $G(k)$.
In this context, it should be noted that Green's function of the Bogoliubov theory still has the $k^{-2}$ dependence for $k\rightarrow 0$
at finite temperatures,\cite{FW72} as both the inverse of the eigenenergy $E_{\bm k}$ and weight factors $(u_{{\bm k}}^2,v_{{\bm k}}^2)$ 
are proportional to $k^{-1}$, which has been known to cause undesirable infrared divergences when used in perturbative calculations.\cite{GN64}

Suitable for confirming the above conjecture may be the functional renormalization group,\cite{Wetterich93,Morris94,Salmhofer99,BTW02,SK06,KBS10,MSHMS12}
which enables us to obtain exact flow equations in terms of an infrared cutoff $\Lambda$
based on microscopic Hamiltonians. 
It has been successfully applied to describe both critical phenomena\cite{KBS10} and low-energy properties of fermions
including ordered phases.\cite{KBS10,MSHMS12} 
The functional formalism has also been applied to Bose-Einstein condensates,\cite{DS07,Wetterich08,FW08,SHK09}
following earlier renormalization-group studies outside it.\cite{CCPS97,PCCS04}
In our viewpoint, however, the renormalization-group approach to Bose-Einstein condensates still has ambiguities about how to choose relevant flow parameters.

To improve this situation, we extend the vertex expansion described in detail by Kopietz {\em et al}.\ \cite{KBS10}
into the Bose-condensed phases.
Specifically, we use a Taylor expansion of the effective action $\Gamma$ 
to derive exact flow equations for the $n$-point vertices $\Gamma^{(n)}$ ($n=0,1,2,\cdots$) as Eq.\ (\ref{Gamma^(n)-eq})
below; the approach was already adopted by Sch\"utz and Kopietz\cite{SK06} so as to be applicable to Bose-Einstein condensates.
In addition, we incorporate into the equations a set of exact identities
obtained by Goldstone's theorem (I), i.e., Eq.\ (\ref{GoldstoneI}) below,
which include
the Hugenholtz-Pines relation\cite{HP59} $\Sigma(0)-\mu=\Delta(0)$ as the lowest-order one
and are also called {\em Ward identities} by Castellani {\em et al}.\cite{CCPS97}
The procedure yields a substantial reduction in the number of flow parameters.
The resulting renormalization-group equations, which are exact and satisfy Goldstone's theorem (I) automatically, 
form a new and reliable basis to study Bose-Einstein condensates theoretically, especially on their low-energy properties.

Using them, we find that (i) the anomalous self-energy $\Delta(0)$ vanishes 
below the critical dimensions $d_{\rm c}=4$ ($d_{\rm c}=3$) at finite temperatures (zero temperature) as Eqs.\ (\ref{dg_Lambda}) and (\ref{dg_Lambda-2})-(\ref{g_Lambda-asymp2}), and (ii)
an exponent (i.e., anomalous dimension) $\eta>0$ is expected to develop in the one-particle density matrix at finite temperatures below $d_{\rm c}=4$.
The exponent $\eta$ will be shown 
to be expressible to the leading order in $\epsilon\equiv 4-d$ as $\eta\propto \epsilon^2$;
see Eqs.\ (\ref{eta^(2b4d)}) and (\ref{eta^(2c0)}) on this point.

It should be noted that the result $\Delta(0)=0$ for $d_{\rm c}\leq 3$ at $T=0$ was already derived by 
 Nepomnyashchi\u{i} and  Nepomnyashchi\u{i},\cite{Nepomnyashchii75,Nepomnyashchii78,Nepomnyashchii83} 
sometimes referred to as  the Nepomnyashchi\u{i} identity,\cite{SHK09}
whose validity has been confirmed by later studies including renormalization-group ones.\cite{Popov79,CCPS97,PCCS04,DS07,SHK09}
On the other hand, the linear dispersion in the one-particle channel has been argued to persist
owing to the vanishing of the frequency renormalization factor $B$,\cite{Nepomnyashchii75}
which is distinct from the Bogoliubov mode,  however.\cite{CCPS97}
Besides reproducing the Nepomnyashchi\u{i} identity at $T=0$, we also show that $\Delta(0)=0$ holds for $d\leq 4$ at $T>0$.
Moreover, our finite-temperature analysis in terms of discrete Matsubara frequencies, where the frequency renormalization factor $B$
becomes irrelevant, indicates clearly that 
the Bogoliubov dispersion should be absent at least at finite temperatures for $d\leq 4$.
Moreover, the argument on the persistence of the linear dispersion by Nepomnyashchi\u{i} and  Nepomnyashchi\u{i}\cite{Nepomnyashchii75,Nepomnyashchii78} 
is implicitly based on the Gavoret-Nozi\`eres result\cite{GN64} that the one- and two-particle Green's functions share a 
common pole, whose validity we have been questioning;\cite{Kita11,TKK16} see also the second paragraph below on this point.
Instead, we here argue that no branch with a linear dispersion exists in the one-particle channel at $T>0$ for $d\leq 4$.
Thus, the present study provides an interpretation of $\Delta(0)=0$ that is completely
different from the one by Nepomnyashchi\u{i} and  Nepomnyashchi\u{i}.
Although they may sound surprising, our finite temperature results will be obtained 
as a direct and natural extension of the Wilson-Fisher $\epsilon$ expansion near the critical point\cite{WF72,KBS10}
into the Bose-condensed phases  based on the exact equations for the vertices, as can be seen by comparing in Sec.\ 4 and Appendix D.
Detailed comparisons with the previous renormalization-group studies\cite{CCPS97,PCCS04,SHK09,DS07,SK06} will be presented in Sec.\ 6.

Penrose and Onsager\cite{PO56,Yang62} introduced the concept of {\em off-diagonal long-range order} (ODLRO) to characterize Bose-Einstein condensation
in terms of the one-particle density matrix $\rho({\bm r})\equiv \langle \hat\psi^\dagger({\bm r}_1)\hat\psi({\bm r}_1+{\bm r})\rangle$  
by $\rho({\bm r})\rightarrow N_{\bm 0}/V>0$ ($r\rightarrow \infty$), where $N_{\bm 0}$ is the number of condensed particles.
However, this concept cannot distinguish interacting Bose-Einstein condensates from ideal ones, especially at finite temperatures
where $N_{\bm 0}$ is smaller than the total particle number in both cases.
To make a possible distinction between them,
we here generalize the concept of ODLRO so as to additionally focus on the approach to $N_{\bm 0}/V$ as 
$\rho({\bm r})\rightarrow N_{\bm 0}/V+C r^{-d+2-\eta}$, where $C$ is a constant. 
The exponent $\eta$ introduced here can be a useful index to characterize interacting Bose-Einstein condensates,
together with the superposition over the number of condensed particles discussed previously.\cite{Kita17,SKK18}

It should be noted finally that the present study clearly indicates that Goldstone's theorem (I)  based on linear infinitesimal transformations 
is different in contents from Goldstone's theorem (II) using commutators of symmetry generators.\cite{GSW62,Weinberg96}
Indeed, it was shown that, when applied to Bose-Einstein condensates, the two distinct proofs are relevant to the poles of one- and two-particle 
Green's functions,\cite{Kita11} respectively, which are not necessarily identical.\cite{TKK16}
The work by Watanabe and Murayama\cite{WM13} on Goldstone's theorem (II) predicts a single Nambu-Goldstone mode in the two-particle channel
with a linear dispersion relation $\propto k$ for $k\rightarrow 0$.
On the other hand, the present study relevant to the one-particle channel predicts no modes with a  linear dispersion 
below $d_{\rm c}=4$ ($d_{\rm c}=3$) at finite temperatures (zero temperature).
This fact exemplifies that the two proofs are not identical, 
contrary to the general understanding where no distinction seems to have been made.\cite{Weinberg96}

This paper is organized as follows. Section \ref{Sec2} derives the effective action $\Gamma$ for Bose-Einstein condensates and
a set of identities with its expansion coefficients $\{\Gamma^{(n)}\}_{n=0}^\infty$ based on Goldstone's theorem (I).
Section \ref{Sec3} obtains exact  renormalization-group equations for $\Gamma^{(n)}$
in such a way as to satisfy Goldstone's theorem (I). 
Section \ref{Sec4} shows that $\Delta(0)$ vanishes below $d_{\rm c}=4$ $(d_{\rm c}=3)$ dimensions at finite temperatures (zero temperatures)
by solving the renormalization-group equation at ${\bm k}={\bm 0}$.
Section \ref{Sec5} studies the exponent $\eta$ at finite temperatures.
Section \ref{Sec6} compares the present approach and results with those of the previous ones.
Section \ref{Sec7} summarizes the paper briefly.
Appendix A enumerates symmetry properties of one-particle Green's functions.
Appendix B studies the derivative of the grand partition function in terms of the infrared cutoff $\Lambda$.
Appendix C presents a proof of an identity obeyed by source terms of the renormalization-group equations.
Appendix D gives a brief derivation of the critical exponent $\eta$ of the O(2) symmetric $\phi^4$ model
within the present formalism based on the Wilson-Fisher $\epsilon$ expansion.

\section{Effective Action for Bose-Einstein Condensates}
\label{Sec2}

\subsection{Grand partition fucntion}

We consider a $d$-dimensional system composed of identical particles with mass $m$ and spin 0 in a box of volume $V$
interacting via a contact potential $g_0\delta({\bm r}-{\bm r}')$.
Adopting the units in which $\hbar=2m=k_{\rm B}=1$ ($k_{\rm B}$:  Boltzmann constant),
our action is given by\cite{Amit,Justin96,NO88}
\begin{align}
S=&\,\int   d\xi\,\bar{\psi}(\xi)\left(\frac{\partial}{\partial\tau}-\bm\nabla^2 -\mu\right)\psi(\xi)
\notag \\
&\, 
+\frac{g_0}{2}\int  d\xi
 \int  d\xi'
\delta(\xi-\xi')\bar{\psi}(\xi)\bar{\psi}(\xi')\psi(\xi')\psi(\xi) .
\label{S-def}
\end{align}
Here 
$\xi\equiv ({\bm r},\tau)$ specifies a space-``time'' point with $0\!\leq\! \tau \!\leq\! \beta\equiv 1/T$
($T$: temperature), 
$\psi$ is the complex bosonic field and $\bar{\psi}$ its conjugate, 
$\mu$ is the chemical potential,
and $g_0$ is the bare coupling constant.
We express $(\psi,\bar\psi)=(\psi_1,\psi_2)$ and expand $\psi_j(\xi)$ ($j=1,2$) as
\begin{align}
\psi_j(\xi)=\frac{1}{V^{1/2}}\sum_{\kappa} \psi_j(\kappa) \, e^{i{\bm k}\cdot{\bm r}-i\omega_\ell \tau}  ,
\label{psi-exp}
\end{align}
where $\kappa\equiv ({\bm k},i\omega_\ell)$ with $\omega_\ell \equiv 2\ell\pi/\beta$ 
$(\ell=0,\pm 1,\cdots)$ denoting the Matsubara frequency.
Substituting Eq.\ (\ref{psi-exp}) into Eq.\ (\ref{S-def}), we obtain
\begin{align}
S = &\,-\beta \sum_{\kappa}\psi_2 (-\kappa)G_0^{-1}(\kappa)\psi_1 (\kappa) 
+ \beta\frac{g_0}{2V}\sum_{\kappa_1\kappa_2\kappa_3\kappa_4}\psi_{2}(\kappa_1)
\notag \\
&\,
\times 
\psi_{2}(\kappa_2)\psi_{1}(\kappa_3)\psi_{1}(\kappa_4) \delta_{\kappa_1+\kappa_2+\kappa_3+\kappa_4,0},
\label{S-k}
\end{align}
where $G_0^{-1}(\kappa)$ is defined by
\begin{align}
G_0^{-1}(\kappa)\equiv i\omega_\ell -k^2+\mu .
\label{G_0^-1}
\end{align}
The transition temperature $T_0$ for $g_0\!=\!0$ is given in the present units by
$T_0=4\pi[N/V\zeta(d/2)]^{2/d}$, where $N$ is the particle number and $\zeta(x)$ is the Riemann $\zeta$ function.

We introduce the grand partition function $Z[J]$ for Eq.\ (\ref{S-k}) 
incorporating extra source functions $J_j(\kappa)$ ($j\!=\!1,2$) as\cite{Amit,Justin96,NO88,dDM64}
\begin{align}
Z[J] \equiv&\, \int D[\psi] \, \exp\Biggl[-S+\sum_{j,\kappa} J_j(\kappa) \psi_j(\kappa) \Biggr]  .
\label{Z_J}
\end{align}
The Taylor expansion of $\ln Z[J]$ in $J_j(\kappa)$ yields
\begin{align}
\ln Z[J] =&\, -\beta\Omega
-\sum_{n=1}^\infty \frac{1}{n!\beta}\sum_{j_1,\kappa_1}\cdots \sum_{j_n,\kappa_n}
\notag \\
&\,\times G^{(n)}_{j_1\cdots j_n}(\kappa_1,\cdots,\kappa_n)
J_{j_1}(\kappa_1)\cdots J_{j_n}(\kappa_n) ,
\label{lnZ-exp}
\end{align}
where $\Omega\equiv -\beta^{-1}\ln Z[0]$ is the grand potential, and $G^{(n)}$ is the $\frac{n}{2}$-particle Green's function defined by
\begin{align}
G^{(n)}_{j_1\cdots j_n}(\kappa_1,\cdots,\kappa_n)=-\beta\left.\frac{\delta^n\ln Z[J]}{\delta J_{j_1}(\kappa_1)\cdots \delta J_{j_n}(\kappa_n)} \right|_{J=0}.
\label{G^(n)}
\end{align}
The factor $\beta$ has been introduced so that $G^{(n)}$ has the dimension of ${\rm E}^{-1}=({\rm ML}^2{\rm T}^{-2})^{-1}$.
Note that $G^{(n)}_{j_1\cdots j_n}(\kappa_1,\cdots,\kappa_n)$ is (i) proportional to
$\delta_{\kappa_1+\cdots+\kappa_n,0}$ owing to the momentum-``energy'' conservation
that holds in any homogeneous equilibrium systems, and also (ii) symmetric with 
respect to the arguments $(j_1\kappa_1,j_2\kappa_2,\cdots,j_n\kappa_n)$ by definition.

\subsection{Effective action}

We perform a Legendre transformation of $\ln Z[J]$ in terms of the derivative
\begin{align}
\phi_j(\kappa)\equiv \frac{\delta \ln Z[J]}{\delta J_j(\kappa)} =\langle \psi_j(\kappa)\rangle
\label{phi-def}
\end{align}
as
\begin{align}
\Gamma[\phi] \equiv \frac{1}{\beta}\left(\sum_{j,\kappa} \phi_j(\kappa)J_j(\kappa)-\ln Z[J] \right) ,
\label{Gamma-def}
\end{align}
where $J=J[\phi]$ on the right-hand side as determined by Eq.\ (\ref{phi-def}).
This $\Gamma[\phi]$, which is called {\em effective action} in relativistic quantum field theory,\cite{BTW02} reduces for
$J\rightarrow 0$ to the grand potential.
The formulation so far has widely been known\cite{Amit,Justin96,NO88,dDM64} and also commonly 
used in the context of the functional renormalization group.\cite{Wetterich93,Morris94,Salmhofer99,BTW02,KBS10,MSHMS12}
Here, we focus on the case with a macroscopic condensate at ${\bm k}={\bm 0}$
characterized by
\begin{align}
\Psi_j\equiv \left.V^{-1/2}\frac{\delta \ln Z[J]}{\delta J_j(0)} \right|_{J=0},
\label{Psi_j}
\end{align}
and derive a set of exact flow equations 
obeyed by  the $n$-point vertices $\Gamma^{(n)}$ $(n=0,1,2,\cdots)$.

To this end, we separate the spontaneous part $\Psi_j$ called {\it condensate wave function} 
from $\phi_j(\kappa)$ in Eq.\ (\ref{phi-def}).
Specifically, we introduce the field $\tilde{\phi}_j(\kappa)$ in terms of $Z[J]$ by
\begin{align}
\tilde{\phi}_j(\kappa)\equiv \frac{\delta \ln Z[J]}{\delta J_j(\kappa)} -\delta_{\kappa 0}\left.\frac{\delta \ln Z[J]}{\delta J_j(0)} \right|_{J=0} .
\label{tphi-def}
\end{align}
which satisfies $\left.\tilde{\phi}_j(\kappa)\right|_{J=0}=0$ by definition.
We also choose the phase of the condensate wave function $\Psi_j$ so that 
$\Psi_1=\Psi_2=(N_{\bm 0}/V)^{1/2}\equiv\Psi $ with $N_{\bm 0}$ denoting the number of condensed particles.
Using them, we can express Eq.\ (\ref{phi-def}) as
\begin{align}
\phi_j(\kappa)=\tilde{\phi}_j(\kappa)+\delta_{\kappa 0} V^{1/2}\Psi .
\label{phi_j-div}
\end{align}
Accordingly, we rewrite $\Gamma[\phi]\rightarrow \Gamma[\Psi,\tilde{\phi}]$ in Eq.\ (\ref{Gamma-exp}) 
and expand it in terms of $\tilde{\phi}$ as
\begin{align}
\Gamma[\Psi,\tilde{\phi}]=&\,\sum_{n=0}^\infty \frac{1}{n! V^{n/2-1}}
\sum_{j_1,\kappa_1}\cdots\sum_{j_n,\kappa_n}
\Gamma^{(n)}_{j_1\cdots j_n}(\kappa_1,\cdots,\kappa_n)
\notag \\
&\,\times \tilde{\phi}_{j_1}(\kappa_1)\cdots \tilde{\phi}_{j_n}(\kappa_n),
\label{Gamma-exp}
\end{align}
where the factor $V^{n/2-1}$ has been introduced to make $\Gamma^{(n)}$ intensive, and the argument $\Psi$ has been
omitted from $\Gamma^{(n)}$ for simplicity.
Note that $\Gamma^{(n)}_{j_1\cdots j_n}(\kappa_1,\cdots,\kappa_n)$ is also proportional to
$\delta_{\kappa_1+\cdots+\kappa_n,0}$, symmetric with 
respect to the arguments $(j_1\kappa_1,j_2\kappa_2,\cdots,j_n\kappa_n)$,
and satisfies 
\begin{align}
\left[\Gamma_{j_1\cdots j_n}^{(n)}(\kappa_1,\cdots,\kappa_n)\right]^*=\Gamma_{3-j_1,\cdots ,3-j_n}(-\kappa_1,\cdots,-\kappa_n),
\end{align}
as can be shown based on $\Gamma^*=\Gamma$ for Eq.\ (\ref{Gamma-exp}) and $\psi_j^*(\kappa)=\psi_{3-j}(-\kappa)$ from Eq.\ (\ref{psi-exp}).
It follows from Eqs.\ (\ref{phi-def}), (\ref{Gamma-def}), and (\ref{phi_j-div}) that 
\begin{align}
J_j(\kappa)= \beta\frac{\delta\Gamma[\Psi,\tilde{\phi}]}{\delta\tilde{\phi_j}(\kappa)}
\label{dGamma/dphi=J}
\end{align}
holds. 
Let us  substitute Eq.\ (\ref{Gamma-exp})  into Eq.\ (\ref{dGamma/dphi=J}), take the limit $\tilde{\phi}\rightarrow 0$ subsequently,
and note that $J$ also vanishes in the limit, as it holds naturally by inverting Eq.\ (\ref{tphi-def}) as $J=J[\tilde{\phi}]$.
We thereby obtain
\begin{align}
0=\Gamma^{(1)}_{j}(\kappa)=\delta_{\kappa 0}\Gamma^{(1)}_{j}(0) ,
\label{Gamma^(1)=0}
\end{align}
where $\delta_{\kappa 0}$ originates from the momentum-``energy'' conservation.
Equation (\ref{Gamma^(1)=0}), which also reads $\delta\Gamma/\delta\Psi\bigr|_{\tilde{\phi}=0}=0$, constitutes the equation to determine $\Psi$.
Substitution of Eqs.\ (\ref{Gamma-exp}) and (\ref{Gamma^(1)=0}) into Eq.\ (\ref{dGamma/dphi=J}) yields
\begin{align}
J_j(\kappa) =&\,\beta\sum_{n=1}^\infty \frac{1}{n!V^{(n-1)/2}}
\sum_{j_1,\kappa_1}\cdots\sum_{j_n,\kappa_n}
\notag\\
&\,\times \Gamma^{(1+n)}_{j j_1\cdots j_n }(\kappa,\kappa_1,\cdots,\kappa_n)
\tilde{\phi}_{j_1}(\kappa_1)\cdots \tilde{\phi}_{j_n}(\kappa_n).
\label{J_i-exp}
\end{align}
Equations (\ref{Gamma-exp}), (\ref{Gamma^(1)=0}), and (\ref{J_i-exp}) forms our starting point to derive the renormalization-group equations
for the vertices.

\subsection{Relation between $\ln Z$ and $\Gamma$}

Noting $J=J[\tilde\phi ]$ in Eq.\ (\ref{dGamma/dphi=J}), 
we can transform the differentiation with respect to $\tilde{\phi}_{j'}(\kappa')$ as\cite{Amit,BTW02,KBS10,MSHMS12}
\begin{align}
\frac{\delta}{\delta \tilde{\phi}_{j'}(\kappa')}=&\,\sum_{j_1,\kappa_1}\frac{\delta J_{j_{1}}(\kappa_1)}{\delta \tilde{\phi}_{j'}(\kappa')}\frac{\delta}{\delta J_{j_{1}}(\kappa_1)}
\notag \\
=&\,\beta\sum_{j_1,\kappa_1}\frac{\delta^2\Gamma[\Psi,\tilde{\phi}]}{\delta \tilde{\phi}_{j'}(\kappa')
\delta \tilde{\phi}_{j_1}(\kappa_1)}\frac{\delta}{\delta J_{j_{1}}(\kappa_1)} .
\label{d/dphi}
\end{align}
Operating Eq.\ (\ref{d/dphi}) on Eq.\ (\ref{phi-def}) yields
\begin{align}
\delta_{j'j}\delta_{\kappa'\kappa}=\beta\sum_{j_1,\kappa_1}\frac{\delta^2\Gamma[\Psi,\tilde{\phi}]}{\delta \tilde{\phi}_{j'}(\kappa')
\delta \tilde{\phi}_{j_1}(\kappa_1)}
\frac{\delta^2 \ln Z[J]}{\delta J_{j_1}(\kappa_1)\delta J_j(\kappa)}.
\end{align}
Introducing the matrix notation\cite{KBS10}
\begin{align}
\left[ \frac{\delta}{\delta\tilde\phi}\otimes\frac{\delta}{\delta\tilde\phi}\right]_{j_1\kappa_1,j_2\kappa_2}\equiv \frac{\delta^2}{\delta\tilde{\phi}_{j_1}(\kappa_1)\delta\tilde{\phi}_{j_2}(\kappa_2)},
\label{Gamma-lnZ}
\end{align}
we can express Eq.\ (\ref{Gamma-lnZ}) concisely as
\begin{align}
\frac{\delta}{\delta\tilde\phi}\otimes\frac{\delta}{\delta\tilde\phi}\Gamma[\Psi,\tilde{\phi}] = \left(\beta\frac{\delta}{\delta J}\otimes\frac{\delta}{\delta J} \ln Z[J] \right)^{-1}.
\label{Gamma-lnZ2}
\end{align}
Taking the limit $\tilde{\phi}\rightarrow 0$ in Eq.\ (\ref{Gamma-lnZ2}) and noting Eqs.\ (\ref{lnZ-exp}) and (\ref{Gamma-exp}), 
we obtain\cite{Amit,BTW02,KBS10,MSHMS12}
\begin{align}
\hat\Gamma^{(2)}=-\hat{G}^{(2)\,-1},
\label{Gamma^(2)-G^(2)}
\end{align}
where $\hat{\Gamma}^{(2)}$ and $\hat{G}^{(2)}$ are matrices whose elements are given by
$\bigl(\hat{\Gamma}^{(2)}\bigr)_{j_1\kappa_1,j_2\kappa_2}=\Gamma^{(2)}_{j_1j_2}(\kappa_1,\kappa_2)$
and $\bigl(\hat{G}^{(2)}\bigr)_{j_1\kappa_1,j_2\kappa_2}=G^{(2)}_{j_1j_2}(\kappa_1,\kappa_2)$, respectively.

It is convenient for later purposes to introduce the matrix\cite{KBS10}
\begin{subequations}
\label{calU}
\begin{align}
\hat{U}[\tilde{\phi}]\equiv \frac{\delta}{\delta\tilde\phi}\otimes\frac{\delta}{\delta\tilde\phi}\Gamma[\Psi,\tilde{\phi}] +\hat{G}^{(2)\,-1} .
\label{calU1}
\end{align}
It follows from Eqs.\ (\ref{Gamma-exp}) and (\ref{Gamma^(2)-G^(2)}) that we can expand $\hat{U}$ as\cite{KBS10}
\begin{align}
\hat{U}[\tilde{\phi}]=&\,\sum_{n=1}^\infty \frac{1}{n!V^{n/2}} \sum_{j_1,\kappa_1}\cdots \sum_{j_n,\kappa_n} \hat\Gamma^{(n+2)}_{j_1\cdots j_n}(\kappa_1,\cdots,\kappa_n)
\notag \\
&\,\times \tilde{\phi}_{j_1}(\kappa_1)\cdots\tilde{\phi}_{j_n}(\kappa_n) ,
\label{calU2}
\end{align}
\end{subequations}
where $\hat\Gamma^{(n+2)}$ is a matrix with the elements
\begin{align}
\left[\hat\Gamma^{(n+2)}_{j_1\cdots j_n}(\kappa_1,\cdots,\kappa_n)\right]_{j\kappa,j'\kappa'}\!\!=
\Gamma^{(n+2)}_{jj'j_1\cdots j_n}(\kappa,\kappa',\kappa_1,\cdots,\kappa_n) .
\label{hatGamma^(n)}
\end{align}
Using Eq.\ (\ref{calU1}), we can express Eq.\ (\ref{Gamma-lnZ2}) alternatively as
\begin{align}
\beta\frac{\delta}{\delta J}\otimes\frac{\delta}{\delta J} \ln Z[J]=&\,-\left(\hat{G}^{(2)\,-1} -\hat{U}\right)^{-1}
\notag \\
=&\,-\hat{G}^{(2)}\sum_{\nu=0}^\infty \left(\hat{U}\hat{G}^{(2)}\right)^\nu .
\label{dlnZ-U}
\end{align}

\subsection{Goldstone's theorem (I)}

We derive a set of identities in terms of the expansion coefficients of 
Eq.\ (\ref{Gamma-exp}) based on the proof of Goldstone's theorem (I).\cite{GSW62,Weinberg96} 
Functional $\Gamma$ is invariant under the global gauge transformation
$({\phi}_1,\phi_2)\rightarrow ({\phi}_1',\phi_2')\equiv (e^{i\chi}\phi_1,e^{-i\chi}\phi_2)$, where $\chi$ is an arbitrary real number.
This invariance is also expressible as $\left.{\delta \Gamma[\Psi',\tilde\phi']}/{\delta\chi}\right|_{\chi=0}=0$.
Noting Eq.\ (\ref{phi_j-div}) and using the chain rule, we can transform the equality into
\begin{align*}
\sum_{j,\kappa}\frac{\delta \Gamma[\Psi,\tilde{\phi}]}{\delta \tilde{\phi}_j(\kappa)}(-1)^{j-1}\left[\tilde{\phi}_j(\kappa) +\delta_{\kappa 0}
V^{1/2}\Psi \right] =0.
\end{align*}
Let us substitute Eq.\ (\ref{Gamma-exp}) into this equality, 
sort the series in powers of $\tilde\phi_j(\kappa)$, 
 and use the fact that $\tilde\phi_j(\kappa)$ can be chosen arbitrarily.
We thereby obtain
\begin{align}
&\,\sum_{\nu=1}^n (-1)^{j_\nu-1} \Gamma^{(n)}_{j_1\cdots j_n}(\kappa_1,\cdots,\kappa_n)
\notag \\
=&\, \Psi \sum_j (-1)^{j} \,\Gamma^{(n+1)}_{ j_1\cdots j_{n}j}(\kappa_1,\cdots,\kappa_{n},0),
\label{GoldstoneI}
\end{align}
for $n=1,2,\cdots$, and also Eq.\ (\ref{Gamma^(1)=0}) for $n=0$.
Setting $n=j_1=1$ in Eq.\ (\ref{GoldstoneI}) reproduces the Hugenholtz-Pines relation\cite{HP59}
\begin{subequations}
\label{HP}
\begin{align}
\Gamma_{11}^{(2)}(0,0)=\Gamma^{(2)}_{12}(0,0).
\label{HP1}
\end{align}
On the other hand, those of $n\geq 2$ relates two kinds of vertices whose external legs differ by one in number.
Specifically, Eq.\ (\ref{GoldstoneI}) for $n=2,3$ yields
\begin{align}
\Gamma_{11}^{(2)}(\kappa_1,\kappa_2)=&\,\frac{\Psi}{2} \left[\Gamma^{(3)}_{112}(\kappa_1,\kappa_2,0)-\Gamma^{(3)}_{111}(\kappa_1,\kappa_2,0)\right],
\label{HP2}
\\
0=&\,\Gamma^{(3)}_{122}(\kappa_1,\kappa_2,0)-\Gamma^{(3)}_{121}(\kappa_1,\kappa_2,0) ,
\label{HP2a}
\end{align}
\begin{align}
&\,\Gamma^{(3)}_{111}(\kappa_1,\kappa_2,\kappa_3)
\notag \\
= &\,\frac{\Psi}{3}\left[\Gamma^{(4)}_{1112}(\kappa_1,\kappa_2,\kappa_3,0)
-\Gamma^{(4)}_{1111}(\kappa_1,\kappa_2,\kappa_3,0) \right],
\label{HP3}
\end{align}
\begin{align}
&\,\Gamma^{(3)}_{112}(\kappa_1,\kappa_2,\kappa_3)
\notag \\
=&\, \Psi\left[
\Gamma^{(4)}_{1122}(\kappa_1,\kappa_2,\kappa_3,0) -\Gamma^{(4)}_{1121}(\kappa_1,\kappa_2,\kappa_3,0)\right].
\label{HP4}
\end{align}
\end{subequations}
Whereas Eq.\ (\ref{HP1}) is widely known,\cite{HP59,GSW62,AGD63,Weinberg96}
much less attention has been paid on Eqs.\ (\ref{HP2})-(\ref{HP4}) in the literature.
However, the identities will be shown to play a crucial role in formulating the functional renormalization group for Bose-Einstein condensates 
having gapless excitations in the one-particle channel in accordance with Goldstone's theorem (I).
It should be noted that the identities of Eq. (\ref{HP}) correspond to the
{\em Ward identities} of Castellani {\em et al}.\ \cite{CCPS97} without the gauge field given in a different basis.

The Bogoliubov theory\cite{Bogoliubov47} corresponds to the approximation $\Gamma^{(n)}\approx \Gamma^{(n{\rm B})}$
with the following non-zero elements,
\begin{subequations}
\label{Gamma^(nB)}
\begin{align}
&\,\Gamma^{(2{\rm B})}_{11}(\kappa_1,\kappa_2) = g_0\Psi^2\delta_{\kappa_1+\kappa_2,0},
\label{Gamma^(2B)_11}
\\
&\, \Gamma^{(2{\rm B})}_{12}(\kappa_1,\kappa_2) =\left(g_0\Psi^2+k_1^2-i\omega_{\ell_1}\right)\delta_{\kappa_1+\kappa_2,0},
\label{Gamma^(2B)_12}
\\
&\,\Gamma^{(3{\rm B})}_{112}(\kappa_1,\kappa_2,\kappa_3)=2g_0\Psi \,\delta_{\kappa_1+\kappa_2+\kappa_3,0},
\\
&\,\Gamma^{(4{\rm B})}_{1122}(\kappa_1,\kappa_2,\kappa_3,\kappa_4)= 2g_0 \,\delta_{\kappa_1+\kappa_2+\kappa_3+\kappa_4,0} ,
\end{align}
\end{subequations}
given in terms of the bare coupling constant $g_0$.

\subsection{One-particle Green's functions}

Of central importance below are one-particle Green's functions $G^{(2)}_{j_1j_2}(\kappa_1,\kappa_2)$.
Since the system is homogeneous and in equilibrium,
we can write it as
\begin{align}
G^{(2)}_{j_1j_2}(\kappa_1,\kappa_2)=G_{j_1j_2}(\kappa_1) \,\delta_{\kappa_1+\kappa_2,0},
\label{G^(2)-G}
\end{align}
where $G_{j_1j_2}(\kappa)$'s make up a $2\times 2$ Nambu matrix 
$\hat{G}=(G_{j_1j_2})$ in the particle-hole space.
It is shown in Appendix\ref{App1} that
$\hat{G}(\kappa)$ is expressible in terms of two functions $(G,F)$ of $k\equiv |{\bm k}|$ and $\omega_\ell$ 
satisfying $F(k,i\omega_\ell)=F^*(k,i\omega_\ell)=F(k,-i\omega_\ell)$ and $G^*(k,i\omega_\ell)=G(k,-i\omega_\ell)$ as
\begin{align}
\hat{G}(\kappa)=\begin{bmatrix}-F(k,i\omega_\ell) & G(k,i\omega_\ell) 
\\
G(k,-i\omega_\ell) & -F(k,i\omega_\ell)
\end{bmatrix} .
\label{hatG}
\end{align}
In the non-interacting limit of $g_0\rightarrow 0$, 
Eq.\ (\ref{hatG}) reduces to\cite{NO88}
\begin{align}
\hat{G}_0(\kappa)=\begin{bmatrix}0 & G_0(k,i\omega_\ell) 
\\
G_0(k,-i\omega_\ell) & 0
\end{bmatrix} ,
\label{hatG_0}
\end{align}
where $G_0(k,i\omega_\ell)$ denotes the inverse of Eq.\ (\ref{G_0^-1}).
It should be noted that we have to incorporate the factor $e^{\pm i\omega_\ell 0_+}$ in summing a single $G(k,\pm i\omega_\ell)$ over $\omega_\ell$ 
to place $\bar{\psi}$ to the left of $\psi$ as in the original ordering of Eq.\ (\ref{S-def}).\cite{Luttinger60}

It follows from Eqs.\ (\ref{Gamma^(2)-G^(2)}) and (\ref{G^(2)-G}) that the elements of  
the matrix $\hat{\Gamma}^{(2)}$ can be written as
\begin{align}
\Gamma^{(2)}_{j_1j_2}(\kappa_1,\kappa_2)=-\delta_{\kappa_1+\kappa_2,0}\left[\hat{G}^{-1}(\kappa_2) \right]_{j_1j_2},
\label{Gamma^(2)-G^-1}
\end{align}
where $\hat{G}^{-1}$ denotes the inverse matrix of Eq.\ (\ref{hatG}).
Let us express $\hat{G}^{-1}$ as 
\begin{align}
\hat{G}^{-1}(\kappa)=\hat{G}_0^{-1}(\kappa)-\hat{\Sigma}(\kappa) ,
\label{G^-1}
\end{align}
with
\begin{align}
\hat{G}_0^{-1}(\kappa)=
\begin{bmatrix}
0 & G_0^{-1}(k,-i\omega_\ell)
\\
G_0^{-1}(k,i\omega_\ell) & 0
\end{bmatrix} ,
\label{hatG_0^-1}
\end{align}
\begin{align}
\hat{\Sigma}(\kappa)=
\begin{bmatrix}
\Delta(k,i\omega_\ell) & \Sigma(k,-i\omega_\ell)
\\
\Sigma(k,i\omega_\ell) & \Delta(k,i\omega_\ell)
\end{bmatrix},
\label{hatSigma}
\end{align}
where $(\Sigma,\Delta)$ should have the symmetries
$\Delta(k,i\omega_\ell)=\Delta^*(k,i\omega_\ell)=\Delta(k,-i\omega_\ell)$ and $\Sigma^*(k,i\omega_\ell)=\Sigma(k,-i\omega_\ell)$
so as to be compatible with Eq.\ (\ref{hatG}).
Now, we substitute Eqs.\ (\ref{hatG_0^-1}) and (\ref{hatSigma}) into Eq.\ (\ref{G^-1}),
invert $\hat{G}^{-1}(\kappa)$ subsequently, and equate the resulting expression with Eq.\ (\ref{hatG}).
We can thereby express $G$ and $F$ in terms of the self-energies $(\Sigma,\Delta)$ as
\begin{subequations}
\begin{align}
G(\kappa)=&\,\frac{-G_0^{-1}(-\kappa)+\Sigma(-\kappa)}{-[G_0^{-1}(\kappa)-\Sigma(\kappa)][G_0^{-1}(-\kappa)-\Sigma(-\kappa)]+\Delta(\kappa)^2} ,
\\
F(\kappa)=&\,\frac{\Delta(\kappa)}{-[G_0^{-1}(\kappa)-\Sigma(\kappa)][G_0^{-1}(-\kappa)-\Sigma(-\kappa)]+\Delta(\kappa)^2},
\end{align}
\end{subequations}
with $\pm\kappa\equiv (k,\pm i\omega_\ell)$.

\subsection{Parametrization of vertices}

It is convenient for later purposes to parametrize the vertices satisfying Eq.\ (\ref{HP}) as
\begin{subequations}
\label{Gamma^(n)-parametrize}
\begin{align}
\Gamma_{11}^{(2)}(\kappa,-\kappa)=&\, g\Psi^2+\delta \Gamma^{(2)}_1(\kappa)  \hspace{3mm}\mbox{with}\hspace{3mm}\delta \Gamma^{(2)}_1(0)= 0,
\label{Gamma^(2)_11-parametrize}
\\
\Gamma_{12}^{(2)}(\kappa,-\kappa)=&\, g\Psi^2+\delta \Gamma^{(2)}_1(\kappa)+\delta \Gamma^{(2)}(\kappa) ,
\end{align}
\begin{align}
\Gamma_{111}^{(3)}(\kappa_1,\kappa_2,\kappa_3)= &\, \Gamma^{(3)}_1(\kappa_1,\kappa_2,\kappa_3) ,
\\
\Gamma_{112}^{(3)}(\kappa_1,\kappa_2,\kappa_3)=&\,\Gamma^{(3)}_1(\kappa_1,\kappa_2,\kappa_3) +
2g\Psi\delta_{\kappa_1+\kappa_2+\kappa_3,0}
\notag \\
&\,+\delta \Gamma^{(3)}(\kappa_1,\kappa_2;\kappa_3) ,
\label{Gamma_112}
\end{align}
\begin{align}
&\,\Gamma_{1111}^{(4)}(\kappa_1,\kappa_2,\kappa_3,\kappa_4)= \Gamma^{(4)}_1(\kappa_1,\kappa_2,\kappa_3,\kappa_4) ,
\end{align}
\begin{align}
&\,\Gamma_{1112}^{(4)}(\kappa_1,\kappa_2,\kappa_3,\kappa_4)
\notag \\
= &\,\Gamma^{(4)}_1(\kappa_1,\kappa_2,\kappa_3,\kappa_4)+ \Gamma^{(4)}_2(\kappa_1,\kappa_2,\kappa_3;\kappa_4),
\end{align}
\begin{align}
&\,\Gamma_{1122}^{(4)}(\kappa_1,\kappa_2,\kappa_3,\kappa_4)
\notag \\
=&\, \Gamma^{(4)}_1(\kappa_1,\kappa_2,\kappa_3,\kappa_4) +
\frac{1}{3}\left[\Gamma^{(4)}_2(\kappa_{1},\kappa_{2},\kappa_{3};\kappa_4)
\right.
\notag \\
&\,
+\Gamma^{(4)}_2(\kappa_{2},\kappa_{3},\kappa_{4};\kappa_1)+\Gamma^{(4)}_2(\kappa_{3},\kappa_{4},\kappa_{1};\kappa_2)
\notag \\
&\, \left.+\Gamma^{(4)}_2(\kappa_{4},\kappa_{1},\kappa_{2};\kappa_3)
\right]+2g\delta_{\kappa_1+\kappa_2+\kappa_3+\kappa_4,0}
\notag \\
&\,+\delta \Gamma^{(4)}(\kappa_1,\kappa_2;\kappa_3,\kappa_4) ,
\label{Gamma_1122}
\end{align}
\end{subequations}
where $g$ is also a parameter, and the symbol ``;'' in an array of arguments
divides them into two groups, each of which is symmetric under any permutation.
Substituting Eq.\ (\ref{Gamma^(n)-parametrize}), we can express Eq.\ (\ref{HP}) alternatively as
\begin{subequations}
\label{HP-parametrize}
\begin{align}
\delta \Gamma^{(2)}(0)=0,
\end{align}
\begin{align}
\delta \Gamma^{(2)}_1(\kappa)= \frac{\Psi}{2} \delta \Gamma^{(3)}(\kappa,-\kappa;0) ,
\end{align}
\begin{align}
\Gamma^{(3)}_1(\kappa_1,\kappa_2,\kappa_3)=\frac{\Psi}{3}\Gamma^{(4)}_2(\kappa_{1},\kappa_{2},\kappa_{3};0),
\end{align}
\begin{align}
&\,\delta \Gamma^{(3)}(\kappa_1,\kappa_2;\kappa_3)
\notag \\
=&\,\Psi\delta \Gamma^{(4)}(\kappa_1,\kappa_2;\kappa_3,0)+
\frac{\Psi}{3}\left[\Gamma^{(4)}_2(\kappa_{2},\kappa_{3},0;\kappa_1)
\right.
\notag \\
&\, \left. \!\!+\Gamma^{(4)}_2(\kappa_{3},\kappa_{1},0;\kappa_2)
-2\Gamma^{(4)}_2(\kappa_{1},\kappa_{2},0;\kappa_3)
\right].
\end{align}
\end{subequations}
It follows from Eqs.\ (\ref{Gamma^(2)_11-parametrize}) and (\ref{HP-parametrize}) that $\delta\Gamma^{(n)}(\kappa_1,\cdots,\kappa_n)$'s
for $n=2,3,4$ vanish
when all $\kappa_j$'s are set equal to zero.
It should also be noted that $\Gamma^{(n)}$'s with $n\geq 5$ are expressible 
without $g$, as can be shown by using Eq.\ (\ref{GoldstoneI}).

\section{Exact Renormalization-Group Equations}
\label{Sec3}

\subsection{Derivation}

Following the standard procedure of the functional renormalization group,\cite{KBS10,MSHMS12} 
we replace $G_0^{-1}(\kappa)$ in  Eq.\ (\ref{S-k}) 
by $G_{0\Lambda}^{-1}(\kappa)$ with some infrared cutoff $\Lambda$.
The formulation of Sec.\ \ref{Sec2} also applies to this case with an additional $\Lambda$ dependence
in every basic function and functional.
Specifically, repeating the procedure of Eqs.\ (\ref{Z_J})-(\ref{phi_j-div}) yields the effective action 
\begin{align}
&\,\Gamma_\Lambda[\Psi_\Lambda,\tilde{\phi}]
\notag \\
=&\,\frac{1}{\beta}\Biggl\{\sum_{j,\kappa} 
\left[\tilde{\phi}_j(\kappa)+\delta_{\kappa 0}V^{1/2}\Psi_\Lambda\right]J_j(\kappa)-\ln Z_\Lambda[J] \Biggr\}.
\label{Gamma_Lambda}
\end{align}
where  $\Psi_\Lambda$ and $\tilde{\phi}_{j}(\kappa)$ ($j=1,2$) are defined similarly as Eqs.\ (\ref{Psi_j}) and (\ref{tphi-def})
in terms of $\ln Z_\Lambda[J]$.
Note that $J_j(\kappa)$ in Eq.\ (\ref{Gamma_Lambda}) also acquires a dependence on $\Lambda$ through the process of 
solving Eq.\ (\ref{phi-def}) for a given set of $\tilde{\phi}$, which is omitted here for simplicity, however.

Let us differentiate Eq.\ (\ref{Gamma_Lambda}) with respect to $\Lambda$, 
where we can omit all the implicit dependences through $J$ owing to Eq.\ (\ref{phi-def}).
We thereby obtain
\begin{align}
\partial_\Lambda\Gamma_\Lambda[\Psi_\Lambda,\tilde{\phi}]=\frac{V^{1/2}}{\beta}\sum_j J_j(0)\,\partial_\Lambda\Psi_\Lambda
-\frac{1}{\beta}\partial_\Lambda \ln Z_\Lambda [J].
\label{dGamma_Lambda}
\end{align}
The derivative $\partial_\Lambda \ln Z_\Lambda[J]$ is expressible in terms of 
the functional $\hat{U}_\Lambda$ in Eq.\ (\ref{calU}) as\cite{KBS10} (see also Appendix\ref{App2} for details)
\begin{align}
\partial_\Lambda\ln Z_\Lambda [J]= \frac{1}{2}{\rm Tr}
\left[ \hat{G}_\Lambda^{(2)\,-1}+\sum_{\nu=0}^\infty \left(\hat{U}_\Lambda \hat{G}^{(2)}_\Lambda\right)^\nu\hat{U}_\Lambda\right]\hat{\dot{G}}^{(2)}_\Lambda,
\label{dlnZ}
\end{align}
where $\hat{\dot{G}}_\Lambda^{(2)}$ is defined by
\begin{align}
\hat{\dot{G}}^{(2)}_\Lambda\equiv \left(\hat{1}-\hat{G}_{0\Lambda}^{(2)}\hat{\Sigma}_\Lambda^{(2)}\right)^{-1}
\left(\partial_\Lambda \hat{G}_{0\Lambda}^{(2)}\right)\left(\hat{1}-\hat{\Sigma}_\Lambda^{(2)}\hat{G}_{0\Lambda}^{(2)}\right)^{-1}
\label{dotG}
\end{align}
with
\begin{align}
\hat{\Sigma}_\Lambda^{(2)}\equiv \hat{G}_{0\Lambda}^{(2)\,-1}-\hat{G}_{\Lambda}^{(2)\,-1} .
\label{Sigma^(2)}
\end{align}
Note that the elements of $\hat{\dot{G}}_{\Lambda}$, $\hat{G}_{0\Lambda }$, and $\hat{\Sigma}_\Lambda$ in Eq.\ (\ref{dotG}) can be written as
\begin{subequations}
\label{G_0Lam^(2)-G_0Lam}
\begin{align}
\dot{G}_{\Lambda,j_1j_2}^{(2)}(\kappa_1,\kappa_2)=&\,\dot{G}_{\Lambda,j_1j_2}(\kappa_1)\,\delta_{\kappa_1+\kappa_2,0} ,
\label{dotG^(2)-dotG}
\\
G_{0\Lambda,j_1j_2}^{(2)}(\kappa_1,\kappa_2)=&\,G_{0\Lambda,j_1j_2}(\kappa_1)\,\delta_{\kappa_1+\kappa_2,0} ,
\label{G_0^(2)-G_0}
\\
\Sigma_{\Lambda,j_1j_2}^{(2)}(\kappa_1,\kappa_2)=&\,\delta_{\kappa_1+\kappa_2,0}\,
\Sigma_{\Lambda,j_1j_2}(\kappa_2),
\label{Sigma^(2)-Sigma}
\end{align}
\end{subequations}
similarly as Eqs.\ (\ref{G^(2)-G}) and (\ref{Gamma^(2)-G^-1}).
Functions $\hat{G}_{0\Lambda}^{(2)\,-1}$ and $\hat{G}_{\Lambda}^{(2)\,-1}$ with the dimension ${\rm E}={\rm ML}^2{\rm T}^{-2}$ are also expressible as
Eq.\ (\ref{Sigma^(2)-Sigma}).

\begin{figure}[t]
\begin{center}
\includegraphics[width=0.95\linewidth]{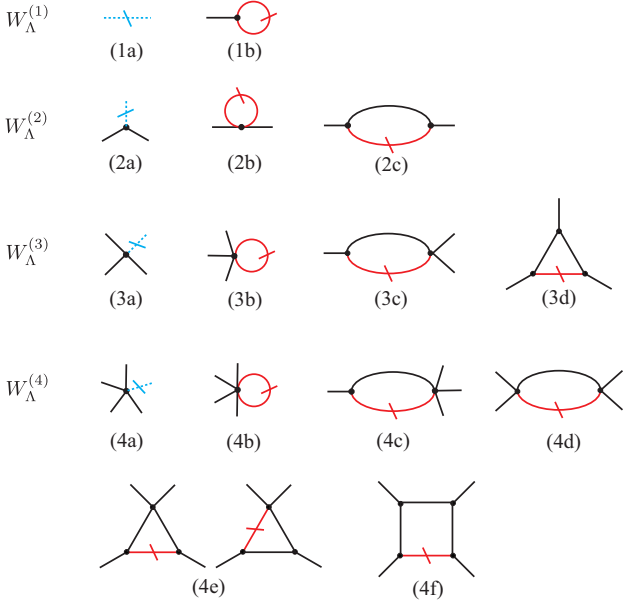}
\end{center}
\caption{Diagrammatic expressions of $W^{(n)}_{\Lambda,j_1\cdots j_n}$ defined by Eq.\ (\ref{W^(n)-def})
for $1\leq n\leq 4$.
A line with a dash (dotted line with a dash) denotes $\dot{G}_{\Lambda,j_1j_2}$ 
($\partial_\Lambda\Psi_\Lambda$).\cite{MSHMS12}
\label{Fig1}}
\end{figure}

Let us substitute Eq.\ (\ref{dlnZ}) into Eq.\ (\ref{dGamma_Lambda}) and expand $\Gamma_\Lambda[\Psi_\Lambda,\tilde{\phi}]$, $J_j(0)$, and $\hat{U}_\Lambda$
in the resulting expression as
Eqs.\ (\ref{Gamma-exp}),  (\ref{J_i-exp}), and (\ref{calU2}).
Subsequently, we equate the coefficient of
${\rm O}(\tilde{\phi}^n)$ for $n=0,1,2,\cdots$ on both sides in a symmetric form with respect to $\tilde{\phi}_j(\kappa)$ using the permutation operator 
\begin{align}
\hat{P}\equiv \left(\begin{array}{ccc} 1 & \cdots & n \\ p_1 & \cdots & p_n\end{array}\right) 
\end{align}
on the right-hand side.
We thereby obtain the exact functional-renormalization-group equations for $\Gamma^{(n)}_\Lambda$ in the form
\begin{align}
-\partial_\Lambda \Gamma_{\Lambda,j_1\cdots j_n}^{(n)}(\kappa_1,\cdots,\kappa_n)=W_{\Lambda,j_1\cdots j_n}^{(n)}(\kappa_1,\cdots,\kappa_n),
\label{Gamma^(n)-eq}
\end{align}
where  $\Gamma^{(1)}_{\Lambda,j}(\kappa)=0$ holds as seen from Eq.\ (\ref{Gamma^(1)=0}).
Functions $W_{\Lambda,j_1\cdots j_n}^{(n)}(\kappa_1,\cdots,\kappa_n)$ for $n=0,1,2,3,4$ are given explicitly 
in terms of Eq.\ (\ref{hatGamma^(n)}) by
\begin{widetext}
\begin{subequations}
\label{W^(n)-def}
\begin{align}
W^{(0)}_{\Lambda}= \frac{1}{2V\beta}{\rm Tr} \,\hat{G}_\Lambda^{(2)\,-1}\hat{\dot G}_\Lambda^{(2)} ,
\label{W^(0)-def}
\end{align}
\begin{align}
W^{(1)}_{\Lambda,j_1}(0)
=  -\sum_{j_2} \Gamma^{(2)}_{\Lambda,j_1j_2}(0,0)\,\partial_\Lambda\Psi_{\Lambda}
+\frac{1}{2V\beta}{\rm Tr} \,
\hat\Gamma^{(3)}_{\Lambda,j_1}(0)\hat{\dot G}_\Lambda^{(2)},
\label{W^(1)-def}
\end{align}
\begin{align}
W^{(2)}_{\Lambda,j_1j_2}(\kappa,-\kappa)
= &\,  -\sum_{j_3} \Gamma^{(3)}_{\Lambda,j_1j_2j_3}(\kappa,-\kappa,0)\,\partial_\Lambda\Psi_{\Lambda}
+  \frac{1}{2V\beta}{\rm Tr} \,
\hat\Gamma^{(4)}_{\Lambda,j_1j_2}(\kappa,-\kappa)\hat{\dot{G}}_\Lambda^{(2)}
+ \frac{1}{2V\beta}{\rm Tr} \biggl[
\hat\Gamma^{(3)}_{\Lambda,j_1}(\kappa)\hat{G}_\Lambda^{(2)}
\hat\Gamma^{(3)}_{\Lambda,j_2}(-\kappa)
\notag \\
&\,+
\hat\Gamma^{(3)}_{\Lambda,j_2}(-\kappa)\hat{G}_\Lambda^{(2)}
\hat\Gamma^{(3)}_{\Lambda,j_1}(\kappa)\biggr]
\hat{\dot{G}}_\Lambda^{(2)} ,
\label{W^(2)-def}
\end{align}
\begin{align}
W^{(3)}_{\Lambda,j_1j_2j_3}(\kappa_1,\kappa_2,\kappa_3)
= &\, - \sum_{j_4} \Gamma^{(4)}_{\Lambda,j_1j_2j_3 j_4}(\kappa_1,\kappa_2,\kappa_3,0)\,\partial_\Lambda\Psi_{\Lambda}
+ \frac{1}{2V\beta}{\rm Tr} \,\hat\Gamma^{(5)}_{\Lambda,j_1 j_2j_3}(\kappa_1,\kappa_2,\kappa_3)\hat{\dot G}_\Lambda^{(2)}
\notag \\
&\, +
\frac{1}{2V\beta} {\rm Tr} \,\frac{1}{2!}\sum_{\hat P} \biggl[\hat\Gamma^{(4)}_{\Lambda,j_{p_1}j_{p_2}}(\kappa_{p_1},\kappa_{p_2})
\hat{G}_\Lambda^{(2)}\hat\Gamma^{(3)}_{\Lambda,j_{p_3}}(\kappa_{p_3})
+\hat\Gamma^{(3)}_{\Lambda,j_{p_1}}(\kappa_{p_1})\hat{G}_\Lambda^{(2)}\hat\Gamma^{(4)}_{\Lambda,j_{p_2}j_{p_3}}(\kappa_{p_2},\kappa_{p_3})\biggr]
\hat{\dot G}_\Lambda^{(2)} 
\notag \\
&\,
+\frac{1}{2V\beta}\sum_{\hat P} {\rm Tr} \,
\hat\Gamma^{(3)}_{\Lambda,j_{p_1}}(\kappa_{p_1})\hat{G}_\Lambda^{(2)}\hat\Gamma^{(3)}_{\Lambda,j_{p_2}}(\kappa_{p_2})\hat{G}_\Lambda^{(2)}\hat\Gamma^{(3)}_{\Lambda,j_{p_3}}(\kappa_{p_3})
\hat{\dot G}_\Lambda^{(2)} ,
\label{W^(3)-def}
\end{align}
\begin{align}
&\, W^{(4)}_{\Lambda,j_1j_2j_3j_4}(\kappa_1,\kappa_2,\kappa_3,\kappa_4)
\notag \\
= &\,-\sum_{j_5} \Gamma^{(5)}_{\Lambda,j_1j_2j_3j_4 j_5}(\kappa_1,\kappa_2,\kappa_3,\kappa_4,0)\,\partial_\Lambda\Psi_{\Lambda}
+ \frac{1}{2V\beta}{\rm Tr} \, 
\hat\Gamma^{(6)}_{\Lambda,j_1 j_2j_3j_4}(\kappa_1,\kappa_2,\kappa_3,\kappa_4)\hat{\dot G}_\Lambda^{(2)}
\notag \\
&\,+\frac{1}{2V\beta}{\rm Tr} \, \frac{1}{3!}\sum_{\hat P} \biggl[\hat\Gamma^{(5)}_{\Lambda,j_{p_1}j_{p_2}j_{p_3}}(\kappa_{p_1},\kappa_{p_2},\kappa_{p_3})\hat{G}_\Lambda^{(2)} \hat\Gamma^{(3)}_{\Lambda,j_{p_4}}(\kappa_{p_4})
+\hat\Gamma^{(3)}_{\Lambda,j_{p_1}}(\kappa_{p_1})\hat{G}_\Lambda^{(2)}\hat\Gamma^{(5)}_{\Lambda,j_{p_2}j_{p_3}j_{p_4}}(\kappa_{p_2},\kappa_{p_3},\kappa_{p_4})\biggr]\hat{\dot G}_\Lambda^{(2)}
\notag \\
&\, +\frac{1}{2V\beta}{\rm Tr} \, \frac{1}{(2!)^2}\sum_{\hat P}\hat\Gamma^{(4)}_{\Lambda,j_{p_1}j_{p_2}}(\kappa_{p_1},\kappa_{p_2})\hat{G}_\Lambda^{(2)} \hat\Gamma^{(4)}_{\Lambda,j_{p_3}j_{p_4}}(\kappa_{p_3},\kappa_{p_4})
\hat{\dot G}_\Lambda^{(2)}
+\frac{1}{2V\beta}{\rm Tr} \, \frac{1}{2!}\!\sum_{\hat P}\biggl[\hat\Gamma^{(4)}_{\Lambda,j_{p_1}j_{p_2}}(\kappa_{p_1},\kappa_{p_2})\hat{G}_\Lambda^{(2)} \hat\Gamma^{(3)}_{\Lambda,j_{p_3}}(\kappa_{p_3})
\notag \\
&\,\times\hat{G}_\Lambda^{(2)} 
 \hat\Gamma^{(3)}_{\Lambda,j_{p_4}}(\kappa_{p_4})
+ \hat\Gamma^{(3)}_{\Lambda,j_{p_1}}\!(\kappa_{p_1})\hat{G}_\Lambda^{(2)}\hat\Gamma^{(4)}_{\Lambda,j_{p_2}j_{p_3}}\!(\kappa_{p_2},\kappa_{p_3})
\hat{G}_\Lambda^{(2)} \hat\Gamma^{(3)}_{\Lambda,j_{p_4}}\!(\kappa_{p_4})
+ \hat\Gamma^{(3)}_{\Lambda,j_{p_1}}(\kappa_{p_1})\hat{G}_\Lambda^{(2)} \hat\Gamma^{(3)}_{\Lambda,j_{p_2}}(\kappa_{p_2})\hat{G}_\Lambda^{(2)}
\notag \\
&\, 
\times \hat\Gamma^{(4)}_{\Lambda,j_{p_3}j_{p_4}}(\kappa_{p_3},\kappa_{p_4})
\biggr]\hat{\dot G}_\Lambda^{(2)}
+\frac{1}{2V\beta}{\rm Tr} \, \sum_{\hat P}
\hat\Gamma^{(3)}_{\Lambda,j_{p_1}}(\kappa_{p_1})\hat{G}_\Lambda^{(2)}\hat\Gamma^{(3)}_{\Lambda,j_{p_2}}(\kappa_{p_2})\hat{G}_\Lambda^{(2)}\hat\Gamma^{(3)}_{\Lambda,j_{p_3}}(\kappa_{p_3})
\hat{G}_\Lambda^{(2)}
 \hat\Gamma^{(3)}_{\Lambda,j_{p_4}}(\kappa_{p_4})\hat{\dot G}_\Lambda^{(2)}  .
\label{W^(4)-def}
\end{align}
\end{subequations}
\end{widetext}
Functions $W_{\Lambda,j_1\cdots j_n}^{(n)}(\kappa_1,\cdots,\kappa_n)$ for $n\geq 5$ can be obtained similarly,
which are not relevant in the present study, however.
Equations (\ref{W^(1)-def})-(\ref{W^(4)-def}) are expressible diagrammatically as
Fig.\ \ref{Fig1}.

It follows from Eq.\ (\ref{Gamma^(1)=0}) that $W^{(1)}_{\Lambda,1}(0)=0$ holds. 
Substituting Eq.\ (\ref{W^(1)-def}) into the equality and noting Eq.\ (\ref{HP1}),
we obtain the renormalization-group equation for the condensate wave function as
\begin{align}
\partial_\Lambda\Psi_{\Lambda} = 
\frac{1}{4V\beta \,\Gamma_{\Lambda,11}^{(2)}(0,0)}{\rm Tr} \,
\hat\Gamma^{(3)}_{\Lambda,1}(0)\,\hat{\dot G}_\Lambda^{(2)} .
\label{dPsi-eq}
\end{align}
Thus, the variation of the condensate wave function is caused by the change in the number of 
excitations given on the right-hand side.

\subsection{Sharp Momentum Cutoff}

The key quantity in the formulation above is $G_{0\Lambda}^{-1}(\kappa)$,
which is constructed here so that $\hat{G}_0(\kappa)$ in Eq.\ (\ref{hatG_0}) is replaced by\cite{KBS10,MSHMS12}
\begin{align}
\hat{G}_{0\Lambda}(\kappa)=\varTheta(k-\Lambda) \hat{G}_0(\kappa),
\label{G_Lambda0}
\end{align}
where $\varTheta(x)$ is the Heaviside step function and $k\equiv |{\bm k}|$.
The corresponding $\hat{G}_{\Lambda}(\kappa)$ and $\hat{\dot{G}}_{\Lambda}(\kappa)$ are expressible as
\begin{subequations}
\label{hatGdG_Lambda}
\begin{align}
\hat{G}_{\Lambda}(\kappa)=&\,\varTheta(k-\Lambda)\hat{\cal G}_\Lambda(\kappa),
\label{hatG_Lambda}
\\
\hspace{5mm}\hat{\dot{G}}_{\Lambda}(\kappa)=&\,-\delta(k-\Lambda) \hat{\cal G}_\Lambda(\kappa) ,
\label{hatdG_Lambda}
\end{align}
\end{subequations}
where  $\hat{\cal G}_\Lambda(\kappa)$ is given in terms of $\hat\Sigma_\Lambda(\kappa)$ in Eq.\ (\ref{Sigma^(2)-Sigma}) by
\begin{align}
\hat{\cal G}_\Lambda(\kappa)\equiv 
\left[\hat{G}_0^{-1}(\kappa)-\hat{\Sigma}_\Lambda(\kappa) \right]^{-1}= -
\left[\hat\Gamma^{(2)}_\Lambda(-\kappa,\kappa)\right]^{-1}  .
\label{calG}
\end{align}
Indeed, Eq.\ (\ref{hatG_Lambda}) is seen to hold by expressing Eq.\ (\ref{Sigma^(2)}) in the form
$\hat{G}_\Lambda^{(2)}=\hat{G}_{0\Lambda}^{(2)}\bigl(\hat{1}-\hat{\Sigma}_\Lambda^{(2)}\hat{G}_{0\Lambda}^{(2)}\bigr)^{-1}$,
expanding it in $\hat{\Sigma}_\Lambda^{(2)}\hat{G}_{0\Lambda}^{(2)}$, and using $[\varTheta(x)]^n=\varTheta(x)$ for $n\geq 1$ and Eq.\ (\ref{G_0Lam^(2)-G_0Lam}).
Similarly, Eq.\ (\ref{hatdG_Lambda}) has been obtained from Eq.\ (\ref{dotG}) by (i)  doubly expanding it in
$\hat{G}_{0\Lambda}^{(2)}\hat{\Sigma}_\Lambda^{(2)}$ and $\hat{\Sigma}_\Lambda^{(2)}\hat{G}_{0\Lambda}^{(2)}$,
(ii) using the equality $\delta(x)[\varTheta(x)]^{n-1}=\delta(x)/n$ for $n\geq 1$ as obtained by differentiating both sides
of $[\varTheta(x)]^n=\varTheta(x)$ in terms of $x$, (iii) also using $\sum_{n_1=0}^\infty\sum_{n_2=0}^\infty x^{n_1+n_2}/(n_1+n_2+1)=
\sum_{n=0}^\infty x^{n}=(1-x)^{-1}$, 
and substituting Eq.\ (\ref{G_0Lam^(2)-G_0Lam}).
The second equality in Eq.\ (\ref{calG}) originates from Eq.\ (\ref{Gamma^(2)-G^-1}).

\subsection{Identities concerning $W_{\Lambda}^{(n)}$}
\label{subsec:3.3}

The exact identities of Eq.\ (\ref{GoldstoneI}) remain valid even in the presence of $\Lambda$,
because the proof is irrelevant to the presence of $\Lambda$.
Differentiating them with respect to $\Lambda$ and using Eq.\ (\ref{Gamma^(n)-eq}),
we also obtain the identities that should be obeyed by $W^{(n)}_\Lambda$'s.
For example, those corresponding to Eqs.\ (\ref{HP1}) and (\ref{HP2}) are given by
\begin{subequations}
\label{W^(23)-identities}
\begin{align}
W^{(2)}_{\Lambda,11}(0,0)=W^{(2)}_{\Lambda,12}(0,0),
\label{W^(2)-identity}
\end{align}
\begin{align}
&\,W^{(2)}_{\Lambda,11}(\kappa,-\kappa)
\notag \\
=&\,-\frac{\Gamma^{(3)}_{\Lambda,112}(\kappa,-\kappa,0)-\Gamma^{(3)}_{\Lambda,111}(\kappa,-\kappa,0)}{2}\partial_\Lambda\Psi_\Lambda 
\notag \\
&\,+\frac{W^{(3)}_{\Lambda,112}(\kappa,-\kappa,0)-W^{(3)}_{\Lambda,111}(\kappa,-\kappa,0)}{2}\Psi_\Lambda .
\label{W^(3)-identity}
\end{align}
\end{subequations}
It is shown in Appendix\ref{App3} that the equalities hold naturally when the basic identities of Eq.\ (\ref{HP}) are fulfilled.

This fact implies that, by starting the integration of Eq.\ (\ref{Gamma^(n)-eq}) at some $\Lambda=\Lambda_0$ with a set of 
vertices satisfying Eq.\ (\ref{GoldstoneI}),
we can automatically obtain its solution for $\Lambda\rightarrow 0$ that is compatible with Goldstone's theorem (I).
A set of such initial vertices can be Eq.\ (\ref{Gamma^(nB)}) by the Bogoliubov approximation.

\section{Vanishing of $\hat{\Gamma}^{(2)}_\Lambda(0,0)$ for $\Lambda\rightarrow 0$}
\label{Sec4}

\subsection{Equation for the coupling constant $g_\Lambda$}

It is shown here that $\hat{\Gamma}^{(2)}_\Lambda(0,0)$ vanishes for $\Lambda\rightarrow 0$ below $d_{\rm c}=4$ ($d_{\rm c}=3$) dimensions
at finite temperatures (zero temperature), contrary to Eqs.\ (\ref{Gamma^(2B)_11}) and (\ref{Gamma^(2B)_12}) of the weak-coupling Bogoliubov theory.\cite{Bogoliubov47}
For this purpose, we can safely omit all the $\kappa$ dependences of the vertices as irrelevant
as in the case of the critical phenomena.\cite{Wilson74,SKMa,Amit,Justin96,KBS10}
Thus, we focus on the vertices $\Gamma^{(n)}_{\Lambda,j_1\cdots j_n}(0,\cdots,0)$,
which can be parametrized as Eq.\ (\ref{Gamma^(n)-parametrize}).
Now, our nonzero elements are given by
\begin{subequations}
\label{Gamma^(n0)}
\begin{align}
&\,\Gamma^{(2)}_{\Lambda,11}(0,0)=\Gamma^{(2)}_{\Lambda,12}(0,0)=g_\Lambda \Psi_\Lambda^2, 
\label{Gamma^(20)}
\\
&\,\Gamma^{(3)}_{\Lambda,112}(0,0,0)=\Gamma^{(3)}_{\Lambda,221}(0,0,0)=2 g_\Lambda \Psi_\Lambda , 
\label{Gamma^(30)}
\\
&\,\Gamma^{(4)}_{\Lambda,1122}(0,0,0,0)=2g_\Lambda ,
\label{Gamma^(40)}
\end{align}
\end{subequations}
having the same form as Eq.\ (\ref{Gamma^(nB)}) of the Bogoliubov theory.
We omit the other vertices in Eq.\ (\ref{Gamma^(n)-parametrize}) as irrelevant.

%
\begin{figure}[t]
\begin{center}
\includegraphics[width=0.95\linewidth]{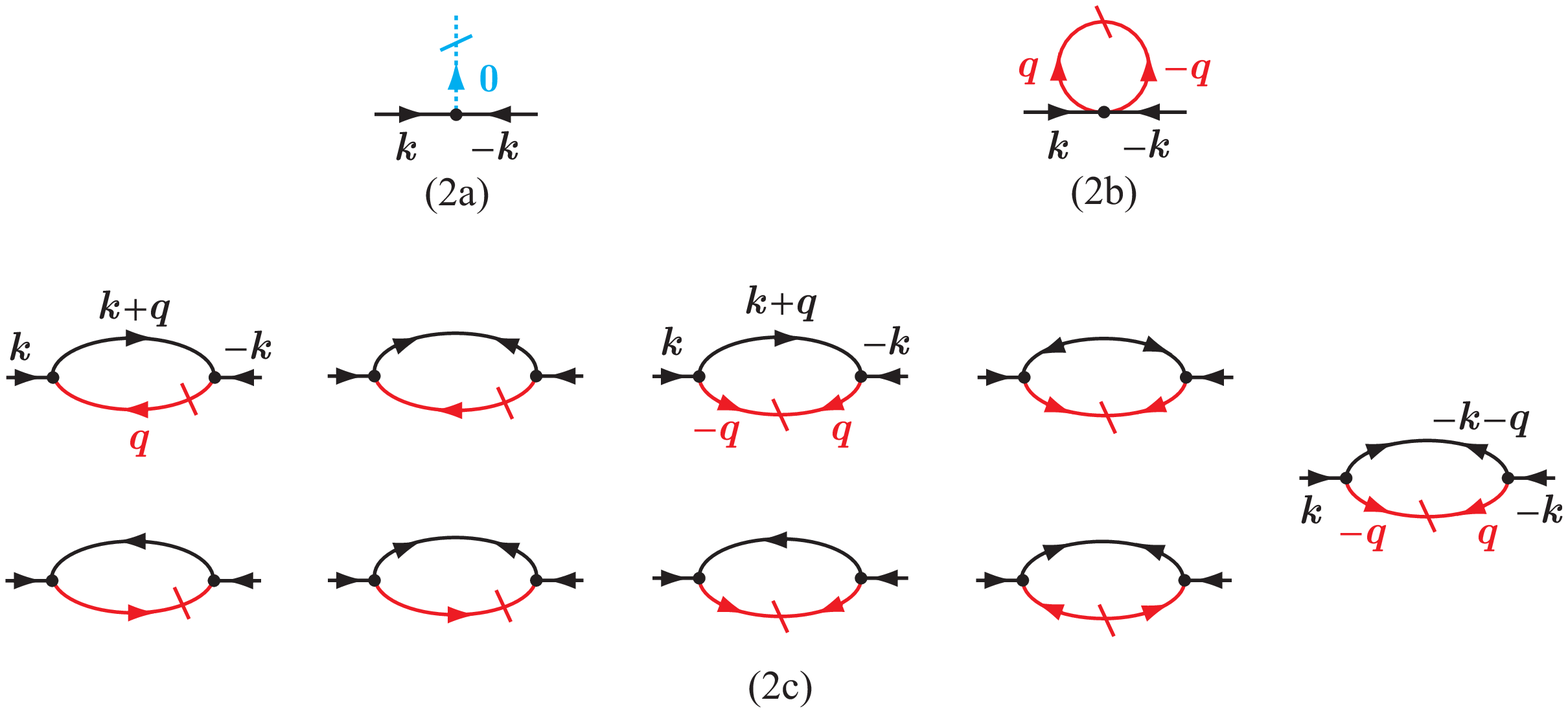}
\end{center}
\caption{Feynman diagrams for $W^{(2)}_{\Lambda,11}$ given by Eq.\ (\ref{W^(2)-def}). 
An incoming (outgoing) arrow at each vertex corresponds to $j=1$ ($j=2$).  Momenta are attached to some of the diagrams for later purposes.
\label{Fig2}}
\end{figure}

Let us substitute Eq.\ (\ref{Gamma^(20)}) into Eq.\ (\ref{dPsi-eq}), 
approximate $\Gamma^{(3)}_{\Lambda,1jj'}(0,-\kappa,\kappa)\approx 
2g_\Lambda\Psi_\Lambda (1-\delta_{j1}\delta_{j'1})$ based on Eq.\ (\ref{Gamma^(30)}),
use the symmetry $G_{21}(\kappa)=G_{12}(-\kappa)$, and adopt the notation of Eq.\ (\ref{hatG}).
We thereby obtain
\begin{align}
\partial_\Lambda\Psi_\Lambda = 
\frac{1}{2V\beta \,\Psi_\Lambda} \sum_\kappa
\left[2\dot{G}_\Lambda(\kappa)\,e^{i\omega_\ell 0_+}-\dot{F}_\Lambda(\kappa)\right] ,
\label{dPsi-eq-2}
\end{align}
where we have also incorporated the factor $e^{i\omega_\ell 0_+}$ following the comment below Eq.\ (\ref{hatG_0}).
Next, we substitute Eq.\ (\ref{Gamma^(n0)})
into Eq.\ (\ref{Gamma^(n)-eq}) for $n=2$, $(j_1,j_2)=(1,1)$, and $\kappa_1=\kappa_2=0$,
i.e., 
$$-\partial_\Lambda \Gamma^{(2)}_{\Lambda,11}(0,0)=W^{(2)}_{\Lambda,11}(0,0).$$
The left-hand side can be transformed as
\begin{subequations}
\begin{align}
-\partial_\Lambda \Gamma^{(2)}_{\Lambda,11}(0,0)=-\Psi_\Lambda^2 \partial_\Lambda g_\Lambda -2g_\Lambda\Psi_\Lambda\partial_\Lambda\Psi_\Lambda .
\label{dGamma^(2)_11(0,0)}
\end{align} 
On the other hand, $W^{(2)}_{\Lambda,11}(0,0)$ in Eq.\ (\ref{W^(2)-def}) is expressible diagrammatically as Fig.\ \ref{Fig2},
which is obtained from Fig.\ \ref{Fig1} (2a), (2b), (2c) 
by adding around each vertex an incoming (outgoing) arrow for the subscript $j=1$ ($j=2$)
so as to be compatible with the approximation of Eq.\ (\ref{Gamma^(n0)}).
The corresponding analytic expression is given by
\begin{align}
&\,W_{\Lambda,11}^{(2)} (0,0)\equiv \lim_{\kappa\rightarrow 0}W_{\Lambda,11}^{(2)} (\kappa,-\kappa)
\notag \\
= &\, -2g_\Lambda\Psi_\Lambda\partial_\Lambda\Psi_\Lambda
-\frac{g_\Lambda}{V \beta}\sum_{\kappa_1} \dot{F}_\Lambda(\kappa_1)
\notag \\
&\,
+  (2g_\Lambda  \Psi_\Lambda)^2 
\Bigl[2\chi_{\Lambda,G\dot{G}}(0)
-2\chi_{\Lambda,F\dot{G}}(0)-2\chi_{\Lambda,G\dot{F}}(0)
\notag \\
&\, +2\chi_{\Lambda,F\dot{F}}(0)+\chi_{\Lambda,F\dot{F}}(0)\Bigr],
\label{W^(2)_11(0,0)}
\end{align}
\end{subequations}
where $\chi_{F\dot{F}}(0)$, for example, is defined by
\begin{align}
\chi_{\Lambda,F\dot{F}}(0)\equiv \lim_{{\bm k}\rightarrow{\bm 0}} \frac{1}{V\beta}
\sum_{\kappa_1}F_\Lambda(|{\bm k}+{\bm k}_1|,i\omega_{\ell_1})\dot{F}_{\Lambda}(k_1,i\omega_{\ell_1}) .
\label{chi}
\end{align}
Equating Eqs.\ (\ref{dGamma^(2)_11(0,0)}) and (\ref{W^(2)_11(0,0)}) yields
\begin{align}
-\partial_\Lambda g_\Lambda =&\, - \frac{g_\Lambda}{V\beta\, \Psi_\Lambda^2 } \sum_\kappa
\dot{F}_\Lambda(\kappa)+  4g_\Lambda^2  \Bigl[2\chi_{\Lambda,G\dot{G}}(0)
\notag \\
&\,
-2\chi_{\Lambda,F\dot{G}}(0)-2\chi_{\Lambda,G\dot{F}}(0)+2\chi_{\Lambda,F\dot{F}}(0)
\notag \\
&\,+\chi_{\Lambda,F\dot{F}}(0)\Bigr] .
\label{dg_Lambda}
\end{align}
This equation, which has been obtained from Eq.\ (\ref{Gamma^(n)-eq}) for $n=2$ and $(j_1,j_2)=(1,1)$, 
will form the basis for clarifying the limiting behavior of $g_\Lambda$ as $\Lambda\rightarrow 0$
 in Sec.\ \ref{subsec4.3} and Sec.\ \ref{subsec4.4}.
The $\chi$ functions in Eq.\ (\ref{dg_Lambda}), which originate from finite $\Gamma^{(3)}$ vertices
characteristic of Bose-Einstein condensation,
will be shown to be responsible for the vanishing of $g_\Lambda$.
Meanwhile, it is also worth pointing out that the same conclusion as obtained below can be reached
by (i) Eq.\ (\ref{Gamma^(n)-eq}) for $n=2$ with $(j_1,j_2)=(1,2)$,
(ii) $n=3$ with $(j_1,j_2,j_3)=(1,1,2)$, and (iii) $n=4$ with $(j_1,j_2,j_3,j_4)=(1,1,2,2)$, 
as it should be according to Eqs.\ (\ref{HP}) and (\ref{W^(23)-identities}).

\subsection{Green's functions}

We focus on the behavior of Eq.\ (\ref{dg_Lambda}) for $\Lambda\rightarrow 0$.
To this end, we approximate $\hat{\Gamma}_\Lambda^{(2)}(-\kappa,\kappa)$ in Eq.\ (\ref{calG})
by noting Eqs.\ (\ref{Gamma^(2)-G^-1})-(\ref{hatSigma}) and (\ref{Gamma^(20)}) as
\begin{align}
&\,\hat{\Gamma}_\Lambda^{(2)}(-\kappa,\kappa) 
\notag \\
=&\, \begin{bmatrix} \Delta_\Lambda(\kappa) & k^2+\Sigma_\Lambda(-\kappa)-\mu+i\omega_\ell \\
k^2+\Sigma_\Lambda(\kappa)-\mu-i\omega_\ell & \Delta_\Lambda(\kappa) \end{bmatrix}
\notag \\
\approx &\,\begin{bmatrix}
\vspace{1mm}
g_\Lambda \Psi_\Lambda^2 +z_{\Lambda,11}^{-1}k^2  & g_\Lambda \Psi_\Lambda^2+z_{\Lambda,12}^{-1}k^2+ i\omega_\ell \\
g_\Lambda \Psi_\Lambda^2+z_{\Lambda,12}^{-1}k^2-i\omega_\ell &  g_\Lambda \Psi_\Lambda^2+z_{\Lambda,11}^{-1}k^2
\end{bmatrix},
\label{Gamma^(2)-approx}
\end{align}
where $z_{\Lambda,j_1j_2}$ is the renormalization factor defined by
\begin{subequations}
\begin{align}
\frac{1}{z_{\Lambda,j_1j_2}}=\left.\frac{\partial {\Gamma}_{\Lambda,j_1j_2}^{(2)}(-\kappa,\kappa)}{\partial k^2}\right|_{\kappa=0}.
\label{z_Lambda,ij}
\end{align}
It is also convenient to introduce the quantities
\begin{align}
z_{\Lambda,\pm}^{-1}\equiv z_{\Lambda,12}^{-1}\pm z_{\Lambda,11}^{-1}= 
\left.\sum_{j=1}^2 (\pm 1)^j\frac{\partial {\Gamma}_{\Lambda,1j}^{(2)}(-\kappa,\kappa)}{\partial k^2}\right|_{\kappa=0}.
\label{z_pm^-1}
\end{align}
\end{subequations}
The function $\hat{\cal G}_\Lambda(\kappa)$ in Eq.\ (\ref{calG}) is then obtained as
\begin{align}
\hat{\cal G}_\Lambda(\kappa)=&\, \frac{1}{i\omega_\ell-E_{\Lambda,k}}
\begin{bmatrix} \vspace{1mm}
- u_{\Lambda,k}v_{\Lambda,k} & u_{\Lambda,k}^2 \\
v_{\Lambda,k}^2 & -u_{\Lambda,k}v_{\Lambda,k}
\end{bmatrix} 
\notag \\
&\,
-\frac{1}{i\omega_\ell+E_{\Lambda,k}}
\begin{bmatrix} \vspace{1mm}
- u_{\Lambda,k}v_{\Lambda,k} & v_{\Lambda,k}^2 \\
u_{\Lambda,k}^2 & -u_{\Lambda,k}v_{\Lambda,k}
\end{bmatrix} ,
\label{G(k->0)}
\end{align}
where $E_{\Lambda,k}$ and $(u_{\Lambda,k},v_{\Lambda,k})$ are 
defined by
\begin{subequations}
\label{Euv}
\begin{align}
E_{\Lambda,k}\equiv &\,k\sqrt{(2g_\Lambda\Psi_\Lambda^2+z_{\Lambda,+}^{-1}k^2)z_{\Lambda,-}^{-1}\,} \,,\\
u_{\Lambda,k}\equiv &\, \sqrt{\frac{1}{2}\left(\frac{g_\Lambda\Psi_\Lambda^2+z_{\Lambda,12}^{-1}k^2}{E_{\Lambda,k}}+1\right)}\,,\\
v_{\Lambda,k}\equiv &\, \sqrt{\frac{1}{2}\left(\frac{g_\Lambda\Psi_\Lambda^2+z_{\Lambda,12}^{-1}k^2}{E_{\Lambda,k}}-1\right)}\,.
\end{align}
\end{subequations}
They are essentially the eigenvalue and eigenvector of the Bogoliubov theory\cite{Bogoliubov47,FW72} 
satisfying $u_{\Lambda,k}^2-v_{\Lambda,k}^2=1$.

The right-hand side of Eq.\ (\ref{dg_Lambda}) for $\Lambda\rightarrow 0$
is dominated by the $\omega_\ell=0$ branch at finite temperatures.\cite{Wilson74,SKMa,Amit,Justin96,KBS10}
For solving the resulting equation, we also invert Eq.\ (\ref{Gamma^(2)-approx}) directly at $\omega_\ell=0$.
We thereby obtain $\hat{\cal G}_\Lambda(k)\equiv \hat{\cal G}_\Lambda(k,i\omega_\ell=0)$ 
valid up to the next-to-leading order in $k^2$ as
\begin{align}
\hat{\cal G}_\Lambda(k)\approx \frac{z_{\Lambda,-}}{2k^2}\begin{bmatrix} 1 & -1 \\ -1 & 1 \end{bmatrix} 
-\frac{1}{4g_\Lambda \Psi_\Lambda^2}\begin{bmatrix} 1 & 1 \\ 1 & 1 \end{bmatrix}  .
\label{calG2}
\end{align}
Thus, $\hat{\cal G}_\Lambda(k)$ at finite temperatures has the same $k^{-2}$ dependence as that 
of the ideal gases to the leading order.

\subsection{Behavior of $g_\Lambda$ for $\Lambda\rightarrow 0$ at finite temperatures}
\label{subsec4.3}

Let us express the matrices in Eq.\ (\ref{hatGdG_Lambda}) as Eq.\ (\ref{hatG}).
It then follows from Eq.\ (\ref{calG2}) that we can approximate
\begin{subequations}
\label{GF-dGF}
\begin{align}
&\,G_\Lambda(k)\approx F_\Lambda(k)\approx -\frac{z_{\Lambda,-}}{2k^2}\varTheta(k-\Lambda) ,
\\
&\,\dot{G}_\Lambda(k)\approx \dot{F}_\Lambda(k)\approx \frac{z_{\Lambda,-}}{2k^2}\delta(k-\Lambda),
\end{align}
\end{subequations}
to the leading order for $k\rightarrow 0$.
Substituting these expressions into Eq.\ (\ref{chi}), omitting all the $\omega_\ell\neq 0$ contributions, 
and using Eq.\ (\ref{GG-lim}), we obtain $\chi_{\Lambda,F\dot{F}}(0)$ at finite temperatures
as
\begin{align}
\chi_{\Lambda,F\dot{F}}(0) \approx &\,-\frac{z_{\Lambda,-}^2}{8\beta}\int\frac{d^d k}{(2\pi)^d}\frac{\delta(k-\Lambda)}{k^4}
\notag \\
=&\,-\frac{z_{\Lambda,-}^2}{8\beta}K_d \Lambda^{-1-\epsilon},
\label{chi(T>0)}
\end{align}
where $\epsilon$ and $K_d$ are defined by
\begin{align}
\epsilon\equiv 4-d,\hspace{10mm} K_d\equiv \frac{S_d}{(2\pi)^d},
\label{epsilon-def}
\end{align}
with $S_d\equiv 2\pi^{d/2}/\varGamma(d/2)$ denoting the area of the unit sphere in $d$ dimensions
given in terms of the Gamma function $\varGamma(x)$.
We substitute Eq.\ (\ref{chi(T>0)}) and $\chi_{\Lambda,G\dot{G}}(0)\approx \chi_{\Lambda,G\dot{F}}(0)\approx \chi_{\Lambda,F\dot{G}}(0)\approx 
\chi_{\Lambda,F\dot{F}}(0)$ into Eq.\ (\ref{dg_Lambda}), and subsequently omit the first term on the right-hand side with no singular behaviors as irrelevant.
The procedure yields the differential equation for $g_\Lambda$ in the infrared limit as
\begin{align}
-\frac{dg_\Lambda}{d\Lambda}= -\frac{z_{\Lambda,-}^2}{2\beta}K_d \Lambda^{-1-\epsilon} g_\Lambda^2 .
\label{dg_Lambda-2}
\end{align}
For $\epsilon\!>\!0$, this equation cannot have a solution that approaches a finite value as $\Lambda\rightarrow 0$.
It follows from Eq.\ (\ref{eta_L-def1}) and Sec.\ \ref{Sec5} that $z_{\Lambda,-}$ behaves for $|\epsilon|\ll 1$ and $\Lambda\rightarrow 0$ as
$z_{\Lambda,-}=\Lambda^\eta$ with $\eta\propto\epsilon^2$.
This fact indicates that the dependence of $z_{\Lambda,-}$ on $\Lambda$ can be neglected in Eq.\ (\ref{dg_Lambda-2})
as compared with $\Lambda^{-1-\epsilon}$.
Integrating Eq.\ (\ref{dg_Lambda-2})  for $0<\epsilon \ll 1$ and $\Lambda\rightarrow 0$, we thereby obtain 
\begin{align}
g_\Lambda= g_*\frac{\beta}{K_d\,z_{\Lambda,-}^2}\Lambda^\epsilon,\hspace{10mm}g_*\equiv 2\epsilon .
\label{g_Lambda-asymp}
\end{align}
Thus, we have shown that $g_\Lambda$ vanishes in the limit $\Lambda\rightarrow 0$  for $d<4$;
the critical dimension $d_{\rm c}$ is 4 at finite temperatures.
This vanishing of $g_\Lambda$ for $d<4$ is similar to the one at the critical point
shown in the pioneering work by Wilson and Fisher,\cite{WF72,Wilson72}
and $g_*$ corresponds the {\em fixed point}
in the critical phenomena.\cite{Wilson74,SKMa,Amit,Justin96,KBS10} 
Indeed, the present formalism can reproduce their results at the critical point, as detailed in Appendix\ref{App4}.
Comparing the derivation of Eq.\ (\ref{g_Lambda-asymp}) with that of Eq.\ (\ref{u_Lambda-sol}) at the transition point, 
we realize a distinct feature in the present case that the vanishing of $g_\Lambda$ is caused cooperatively by the 
connection between different kinds of vertices given by Eq.\ (\ref{HP}),
which is characteristic of Bose-Einstein condensation.

\subsection{Behavior of $g_\Lambda$ for $\Lambda\rightarrow 0$ at zero temperature}
\label{subsec4.4}

Next, we focus on Eq.\ (\ref{dg_Lambda}) at $T=0$.
We can calculate the key quantity, Eq.\ (\ref{chi}) at $T=0$,  by (i) substituting the (1,1) element of Eq.\ (\ref{G(k->0)}), 
(ii) performing the sum over $\omega_\ell$ 
using the Bose distribution function $f(\varepsilon)\equiv (e^{\,\beta\varepsilon}-1)^{-1}$,\cite{FW72}
and (ii) taking the limit $T\rightarrow 0$ subsequently.
We thereby obtain
\begin{align}
&\,\chi_{\Lambda,F\dot{F}}(0)
\notag \\
=&\,-\frac{1}{V}\sum_{\bm k}(u_{\Lambda,k}v_{\Lambda,k})^2 \left[-f'(E_{\Lambda,k})+\frac{1+2f(E_{\Lambda,k})}{2E_{\Lambda,k}}\right]
\notag \\
&\,\times \delta(k-\Lambda)
\notag \\
\stackrel{T\rightarrow 0}{\longrightarrow}&\,
-\int\frac{d^d k}{(2\pi)^d} \delta(k-\Lambda)\frac{(u_{\Lambda,k}v_{\Lambda,k})^2}{2E_k}
\notag \\
=&\,-\frac{z_{\Lambda,-}^{3/2}\Psi_\Lambda K_d}{16\sqrt{2}} g_\Lambda^{1/2}\Lambda^{d-4}.
\label{chi(T=0)}
\end{align}
Similar calculations yield  $\chi_{\Lambda,G\dot{G}}(0)\approx \chi_{\Lambda,G\dot{F}}(0)\approx \chi_{\Lambda,F\dot{G}}(0)\approx 
\chi_{\Lambda,F\dot{F}}(0)$ to the leading order.
Substituting them into Eq.\ (\ref{dg_Lambda}), one can show that $g_{\Lambda}$ vanishes for $\Lambda\rightarrow 0$ below the critical dimensions, 
$d_{\rm c}=3$. Especially, the limiting behavior of $g_{\Lambda}$ at $d=d_{\rm c}$ is given by
\begin{align}
g_\Lambda = \left(-\frac{3z_{\Lambda,-}^{3/2}\Psi_\Lambda}{8\sqrt{2}} K_3\ln\Lambda \right)^{-2/3} .
\label{g_Lambda-asymp2}
\end{align}
Thus, we have reproduced the Nepomnyashchi\u{i} identity at $T=0$.\cite{Nepomnyashchii75,Nepomnyashchii78}

\section{Anomalous Dimension}
\label{Sec5}

It is shown here for $d\lesssim d_{\rm c}=4$ at finite temperatures 
that the vanishing of the interaction as Eq.\ (\ref{g_Lambda-asymp}) also accompanies 
development of the anomalous dimension
in Green's function $\hat{G}_\Lambda(k)$.

\subsection{Definition}
\label{Sec5.1}

It is well known in the theory of critical phenomena\cite{Wilson74,SKMa,Amit,Justin96,KBS10} that the renormalization factor $z_\Lambda$ 
of Green's function $G_\Lambda(k)\approx z_\Lambda/k^2$ also vanishes for $\Lambda\rightarrow 0$
at the critical point; see also Appendix\ref{App4} on this point.
Looking at Eq.\ (\ref{calG2}),
it is natural to expect that our renormalization factor $z_{\Lambda,-}$ may also vanish for $\Lambda\rightarrow 0$.
To confirm the conjecture, we introduce the {\it flowing anomalous dimension}\cite{KBS10} by
\begin{subequations}
\label{eta_L-def}
\begin{align}
\eta_\Lambda \equiv &\,-\Lambda\,\partial_\Lambda \ln z_{\Lambda,-}^{-1}.
\label{eta_L-def1}
\end{align}
This definition implies assuming the limiting behavior of the renormalization factor as
$z_{\Lambda,-} \sim\Lambda^{\eta_\Lambda}$. 
Let us substitute Eq.\ (\ref{z_pm^-1}) into Eq.\ (\ref{eta_L-def1}), subsequently use Eq.\ (\ref{Gamma^(n)-eq}), 
and omit the $\omega_\ell\neq 0$ contribution as irrelevant at finite temperatures. 
We can thereby transform Eq.\ (\ref{eta_L-def1}) into
\begin{align}
\eta_\Lambda = &\left. \Lambda\, z_{\Lambda,-}  \frac{\partial \delta W_{\Lambda}^{(2)}(k)}{\partial k^2}
\right|_{k^2=0}= \left.\frac{\Lambda\, z_{\Lambda,-} }{2} \frac{\partial^2 \delta W_{\Lambda}^{(2)}(k)}{\partial k^2}
\right|_{k=0} ,
\label{eta_L-def2}
\end{align}
\end{subequations}
with
\begin{subequations}
\label{dW^(2)}
\begin{align}
\delta W_{\Lambda}^{(2)}(k)\equiv \sum_{j=1}^2(-1)^j W_{\Lambda,1j}^{(2)}(-{\bm k},{\bm k}) ,
\label{dW^(2)-1}
\end{align}
where we have removed $\omega_\ell=0$ from the arguments on the right-hand side.
This function is expressible as a sum of the three terms
corresponding to Fig.\ 1 (2a), (2b), (2c) as
\begin{align}
\delta W_{\Lambda}^{(2)}(k)=&\, \sum_{\alpha={\rm a},{\rm b},{\rm c}}\delta W_{\Lambda}^{(2\alpha)}(k) .
\label{dW^(2)-2}
\end{align}
\end{subequations}
For example, $\delta W_{\Lambda}^{(2{\rm b})}(k)$
is obtained from the second term on the right-hand side of Eq.\ (\ref{W^(2)-def}) as
\begin{align*}
&\, \delta W_{\Lambda}^{(2{\rm b})}(k)
\notag \\
=&\, \frac{1}{2\beta}\int\frac{d^dq}{(2\pi)^d}
\sum_{j j_1j_2}(-1)^j \,\Gamma^{(4)}_{\Lambda,1jj_2j_1}(-{\bm k},{\bm k},-{\bm q},{\bm q})
\notag \\
&\,\times \dot{G}_{\Lambda,j_1j_2} (q).
\end{align*}
This expression indicates that we need to know the ${\bm k}$ dependence of the interaction vertices
for calculating Eq.\ (\ref{eta_L-def2}).
Note that Eq.\ (\ref{dW^(2)}) satisfies $\delta W_{\Lambda}^{(2)}(0)=0$ as a whole, as shown in Appendix\ref{App3}.

The  exponent $\eta$ is defined in terms of Eq.\ (\ref{eta_L-def}) by
\begin{align}
\eta\equiv \lim_{\Lambda\rightarrow 0} \eta_\Lambda = \left.\lim_{\Lambda\rightarrow 0} \sum_{\alpha={\rm a},{\rm b},{\rm c}}\frac{\Lambda\, z_{\Lambda,-}}{2}  \frac{\partial^2 \delta W_{\Lambda}^{(2\alpha)}(k)}{\partial k^2}
\right|_{k=0}.
\label{eta-def}
\end{align}

\subsection{Relevant vertices}
\label{Sec5.2}

As the interaction vertices relevant to $\eta$,
we consider only $\delta \Gamma^{(3)}_\Lambda$ and $\delta \Gamma^{(4)}_\Lambda$
on the right-hand sides of Eqs.\ (\ref{Gamma_112}) and (\ref{Gamma_1122}), 
in accordance with the approximation of Eq.\ (\ref{Gamma^(n0)}) where $(\Gamma^{(3)}_{\Lambda,1},\Gamma^{(4)}_{\Lambda,1},\Gamma^{(4)}_{\Lambda,2})\equiv (\Gamma^{(3)}_{\Lambda,111},\Gamma^{(4)}_{\Lambda,1111},\Gamma^{(4)}_{\Lambda,1112}-\Gamma^{(4)}_{\Lambda,1111})$ were neglected even at zero momenta. 
Specifically, our finite vertices are given by
\begin{subequations}
\label{Gamma^(n)-approx}
\begin{align}
\Gamma^{(3)}_{\Lambda,112}({\bm k}_1,{\bm k}_2,{\bm k}_3)
= &\, 2g_\Lambda\Psi_\Lambda\delta_{{\bm k}_1+{\bm k}_2+{\bm k}_3,{\bm 0}}
\notag \\
&\,+\delta\Gamma^{(3)}_\Lambda({\bm k}_1,{\bm k}_2;{\bm k}_3),
\\
\Gamma^{(4)}_{\Lambda,1122}({\bm k}_1,{\bm k}_2,{\bm k}_3,{\bm k}_4)= &\, 2g_\Lambda\delta_{{\bm k}_1+{\bm k}_2+{\bm k}_3+{\bm k}_4,{\bm 0}}
\notag \\
&\, +\delta\Gamma^{(4)}_\Lambda({\bm k}_1,{\bm k}_2;{\bm k}_3,{\bm k}_4),
\end{align}
\end{subequations}
with 
\begin{align}
\delta \Gamma^{(3)}_\Lambda({\bm 0},{\bm 0};{\bm 0})=\delta \Gamma^{(4)}_\Lambda({\bm 0},{\bm 0};{\bm 0},{\bm 0})=0
\label{dGamma^(3,4)(0)=0}
\end{align}
by definition according to the comment below Eq.\ (\ref{Gamma^(n)-parametrize}).

The equations for determining $\delta \Gamma^{(n)}_\Lambda$ ($n=3,4$) are obtained as follows.
Let us solve Eq.\ (\ref{Gamma^(n)-parametrize}) to express $\delta \Gamma^{(n)}_\Lambda$ in terms of $\Gamma^{(n)}_{\Lambda,j_1\cdots j_n}$,
differentiate the resulting $\delta \Gamma^{(n)}_\Lambda$ with respect to $\Lambda$, 
and substitute Eq.\ (\ref{Gamma^(n)-eq}).
We thereby find that $\delta \Gamma^{(n)}_\Lambda$'s exactly obey the equations,
\begin{align}
-\partial_\Lambda\delta\Gamma^{(n)}_\Lambda({\bm k}_1,\cdots,{\bm k}_n)=
\delta W^{(n)}_\Lambda({\bm k}_1,\cdots,{\bm k}_n) ,
\label{dGamma^(n)-eq}
\end{align}
where $\delta W^{(3)}_\Lambda$ and $\delta W^{(4)}_\Lambda$ are defined by
\begin{subequations}
\label{dGamma-dGamma}
\begin{align}
&\,\delta W^{(3)}_\Lambda({\bm k}_1,{\bm k}_2,{\bm k}_3)
\notag \\
=&\,W^{(3)}_{\Lambda,112}({\bm k}_1,{\bm k}_2,{\bm k}_3)-W^{(3)}_{\Lambda,111}({\bm k}_1,{\bm k}_2,{\bm k}_3),
\label{deltaW^(3)}
\end{align}
\begin{align}
&\,\delta W^{(4)}_\Lambda({\bm k}_1,{\bm k}_2,{\bm k}_3,{\bm k}_4)
\notag \\
=&\,W^{(4)}_{\Lambda,1122}({\bm k}_1,{\bm k}_2,{\bm k}_3,{\bm k}_4)
-\frac{1}{3}\biggl[W^{(4)}_{\Lambda,1112}({\bm k}_1,{\bm k}_2,{\bm k}_3,{\bm k}_4)
\notag \\
&\,+W^{(4)}_{\Lambda,1112}({\bm k}_2,{\bm k}_3,{\bm k}_4,{\bm k}_1)
+W^{(4)}_{\Lambda,1112}({\bm k}_3,{\bm k}_4,{\bm k}_1,{\bm k}_2)
\notag \\
&\,+W^{(4)}_{\Lambda,1112}({\bm k}_4,{\bm k}_1,{\bm k}_2,{\bm k}_3)
-W^{(4)}_{\Lambda,1111}({\bm k}_1,{\bm k}_2,{\bm k}_3,{\bm k}_4)\biggr] ,
\label{deltaW^(4)}
\end{align}
\end{subequations}
satisfying
\begin{align}
\delta W^{(3)}_\Lambda({\bm 0},{\bm 0};{\bm 0})=\delta W^{(4)}_\Lambda({\bm 0},{\bm 0};{\bm 0},{\bm 0})=0,
\label{dW^(3,4)(0)=0}
\end{align}
as seen from Eqs.\ (\ref{dGamma^(3,4)(0)=0}) and (\ref{dGamma^(n)-eq}).
We can calculate Eq.\ (\ref{dGamma-dGamma}) based on Eqs.\ (\ref{W^(3)-def}) and (\ref{W^(4)-def}) by
adopting the approximation of Eq.\ (\ref{Gamma^(n0)}) for the vertices.
The resulting $(\delta W^{(3)}_\Lambda,\delta W^{(4)}_\Lambda)$ are substituted into Eq.\ (\ref{dGamma^(n)-eq})
to obtain $(\delta \Gamma^{(3)}_\Lambda,\delta \Gamma^{(4)}_\Lambda)$ by integration.

\begin{figure}[t]
\begin{center}
\includegraphics[width=0.95\linewidth]{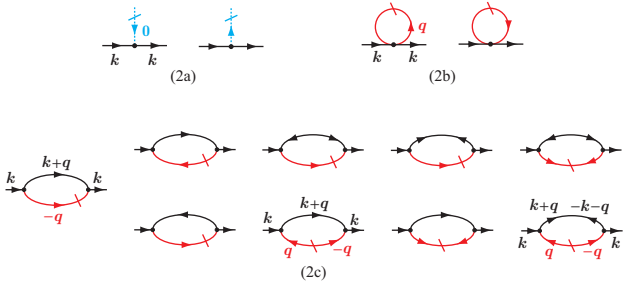}
\end{center}
\caption{Feynman diagrams for $W^{(2)}_{\Lambda,12}$
within the approximation of Eq.\ (\ref{Gamma^(n)-approx}).
An incoming (outgoing) arrow at each vertex corresponds to $j=1$ ($j=2$).
\label{Fig3}}
\end{figure}

The improved vertices of Eq.\ (\ref{Gamma^(n)-approx}) are then used in Eq.\ (\ref{W^(2)-def}) for 
$W^{(2)}_{11}({\bm k},-{\bm k})$ and  $W^{(2)}_{12}({\bm k},-{\bm k})$, which are expressible
diagrammatically as Figs.\ \ref{Fig2} and \ref{Fig3}, respectively.
Also using Eq.\ (\ref{dPsi-eq-2}), we 
obtain analytic expressions for the three contributions of Eq.\ (\ref{dW^(2)-2}) in terms of $\delta \Gamma^{(n)}_\Lambda$'s as
\begin{subequations}
\label{W^(2)-2}
\begin{align}
\delta W_{\Lambda}^{(2{\rm a})}(k)
=&\, -\frac{2\delta\Gamma^{(3)}_{\Lambda}({\bm k},{\bm 0};-{\bm k})-\delta\Gamma^{(3)}_{\Lambda}({\bm k},-{\bm k};{\bm 0})}{2\beta\,\Psi_\Lambda} 
\notag \\
&\,
\times \int\frac{d^dq}{(2\pi)^d}\left[ 2\dot{G}_{\Lambda} (q)-\dot{F}_{\Lambda} (q)\right],
\label{W^(2a)-2}
\end{align}
\begin{align}
\delta W_{\Lambda}^{(2{\rm b})}(k)
=&\, \frac{1}{2\beta} \int\frac{d^dq}{(2\pi)^d}
\left[2\delta\Gamma^{(4)}_{\Lambda}({\bm k},{\bm q};-{\bm k},-{\bm q})  \dot{G}_{\Lambda} (q)
\right.
\notag \\
&\,\left.
+\delta\Gamma^{(4)}_{\Lambda}({\bm k},-{\bm k};{\bm q},-{\bm q})  \dot{F}_{\Lambda} (q)\right],
\label{W^(2b)-2}
\end{align}
with $\dot{G}_\Lambda\equiv\dot{G}_{\Lambda,12}$ and  $\dot{F}_\Lambda\equiv -\dot{G}_{\Lambda,11}$, and
\begin{align}
&\,\delta W^{(2{\rm c})}_\Lambda(k)
\notag \\
=&\,\frac{2g_\Lambda}{\beta}\!\int\!\frac{d^d q}{(2\pi)^d}\!
\Bigl[F_{\Lambda}(|{\bm k}\!+\!{\bm q}|)\delta(q-\Lambda)-\varTheta(|{\bm k}\!+\!{\bm q}|-\Lambda)
\notag \\
&\, \times \dot{F}_{\Lambda}(q)\Bigr]+\frac{2}{\beta \Psi_\Lambda}\int\frac{d^d q}{(2\pi)^d}
\left\{ \delta \Gamma^{(3)}_{\Lambda}({\bm q},-{\bm k}-{\bm q};{\bm k})
\right.
\notag \\
&\,\times
\left[F_{\Lambda}(|{\bm k}+{\bm q}|)\delta(q-\Lambda)-\varTheta(|{\bm k}+{\bm q}|-\Lambda)\dot{F}_{\Lambda}(q)\right]
\notag \\
&\, + \left[\delta \Gamma^{(3)}_{\Lambda}({\bm k},{\bm q};-{\bm k}-{\bm q})-\delta \Gamma^{(3)}_{\Lambda}({\bm k},-{\bm k}-{\bm q};{\bm q})
\right]
\notag \\
&\,\left.\times
\left[\varTheta(|{\bm k}+{\bm q}|-\Lambda)\dot{F}_{\Lambda}(q)+F_{\Lambda}(|{\bm k}+{\bm q}|)\delta(q-\Lambda)\right]\right\} ,
\label{W^(2c)-2}
\end{align}
\end{subequations}
with  ${F}_\Lambda\equiv -{G}_{\Lambda,11}$.
See Appendix\ref{App5} for the derivation of Eq.\ (\ref{W^(2c)-2}).

\subsection{Rescaling}
\label{Sec5.3}

It is useful for studying Eq.\ (\ref{eta-def}) to introduce the dimensionless quantities,
\begin{subequations}
\label{tildes}
\begin{align}
&\, \tilde{\bm k}\equiv \frac{{\bm k}}{\Lambda},\hspace{10mm}x\equiv -\ln \frac{\Lambda}{\Lambda_0}, 
\label{tildek-def}
\end{align}
and express $\delta W_{\Lambda}^{(2)}$, $(\delta {\Gamma}^{(n)}_{\Lambda},\delta W^{(n)}_{\Lambda})$ for $n=3,4$, and
$(\hat{G}_\Lambda,\hat{\dot{G}}_\Lambda)$ as
\begin{align}
&\, \delta W_{\Lambda}^{(2)}({\bm k}) 
=\frac{\Lambda}{z_{\Lambda,-}}\sum_{\alpha={\rm a},{\rm b},{\rm c}}\delta\tilde{W}^{(2\alpha)}_{x}(\tilde{k}),
\label{tildeW^(2)-def}
\end{align}
\begin{align}
&\,
\delta {\Gamma}^{(n)}_{\Lambda}({\bm k}_1,\cdots,{\bm k}_n) = \frac{\Lambda^{4-d}}{z_{\Lambda,-}^2}\delta \tilde{\Gamma}^{(n)}_{x}(\tilde{\bm k}_1,\cdots,\tilde{\bm k}_n),
\label{tildeGamma^(n)-def}
\end{align}
\begin{align}
&\, 
\delta W^{(n)}_{\Lambda}({\bm k}_1,\cdots,{\bm k}_n) = \frac{\Lambda^{3-d}}{z_{\Lambda,-}^2} \delta \tilde{W}^{(n)}_{x}(\tilde{\bm k}_1,\cdots,\tilde{\bm k}_n),
\label{tildeW^(n)-def}
\end{align}
\begin{align}
&\, \hat{G}_\Lambda(k)=\frac{z_{\Lambda,-}}{\Lambda^2}\hat{\tilde{G}}(\tilde{k}),\hspace{10mm}
\hat{\dot{G}}_\Lambda(k)=\frac{z_{\Lambda,-}}{\Lambda^3}\hat{\dot{\tilde{G}}}(\tilde{k}).
\label{tildeG-def}
\end{align}
\end{subequations}
Note that Eq.\ (\ref{tildeGamma^(n)-def}) is different from the conventional rescaling for the critical phenomena;\cite{Amit,Justin96,KBS10}
the rational for the choice is that $\tilde{\Gamma}^{(n)}_x$'s ($n=3,4$) acquire the same physical dimensions as
$g_*$ in Eq.\ (\ref{g_Lambda-asymp}) in terms of the renormalization factors $(\Lambda,z_{\Lambda,-})$.
It follows from Eqs.\ (\ref{hatGdG_Lambda}), (\ref{calG2}), and (\ref{tildeG-def}) that $\hat{\tilde{G}}(\tilde{k})$ 
and $\hat{\dot{\tilde{G}}}(\tilde{k})$ are given to the leading order by
\begin{subequations}
\begin{align}
\hat{\tilde{G}}(\tilde{k})=&\,\frac{ \varTheta(\tilde{k}-1)}{2\tilde{k}^2}\begin{bmatrix} 1 & -1 \\ -1 & 1 \end{bmatrix}, 
\\
\hat{\dot{\tilde{G}}}(\tilde{k})=&\,-\frac{ \delta(\tilde{k}-1)}{2\tilde{k}^2}\begin{bmatrix} 1 & -1 \\ -1 & 1 \end{bmatrix}.
\label{tG(k->0)-T>0}
\end{align}
\end{subequations}
Using Eqs.\ (\ref{tildek-def}) and (\ref{tildeW^(2)-def}), we can express Eq.\ (\ref{eta-def}) concisely as
\begin{align}
\eta =\left. \frac{1}{2}\sum_{\alpha={\rm a},{\rm b},{\rm c}}
\frac{\partial^2 \delta\tilde{W}_{\infty}^{(2\alpha)}(\tilde{k})}{\partial \tilde{k}^2}\right|_{\tilde{k}=0}.
\label{eta-def2}
\end{align}
On the other hand, Eq.\ (\ref{dGamma^(n)-eq}) for $n=3,4$ and Eq.\ (\ref{eta_L-def1}) can be integrated formally 
\begin{subequations}
\begin{align}
\delta\Gamma_\Lambda^{(n)}({\bm k}_1,\cdots,{\bm k}_n)=&\, \int_\Lambda^{\Lambda_0}d\Lambda' \delta W_{\Lambda'}^{(n)}({\bm k}_1,\cdots,{\bm k}_n) ,
\label{dGamma^(n)}
\\
z_{\Lambda,-}=&\, \exp\left[-\int_\Lambda^{\Lambda_0}d\Lambda'\frac{\eta_{\Lambda'}}{\Lambda'}\right] ,
\label{z_L}
\end{align}
\end{subequations}
where we have used the initial condition $\delta \Gamma^{(n)}_{\Lambda_0}({\bm k}_1,\cdots,{\bm k}_n)=0$
compatible with Eq.\ (\ref{Gamma^(n0)}) at $\Lambda=\Lambda_0$.
Let us perform the transformation of Eq.\ (\ref{tildes}) in Eq.\ (\ref{dGamma^(n)}), substitute Eq.\ (\ref{z_L}),
and make a change of variables $x'=x-y$. 
We can thereby express Eq.\ (\ref{dGamma^(n)}) alternatively in terms of $\epsilon\equiv 4-d$ and $\tilde{\eta}_x\equiv \eta_\Lambda$ as
\begin{align}
\delta \tilde{\Gamma}^{(n)}_{x}(\tilde{\bm k}_1,\cdots,\tilde{\bm k}_n)
= &\,\int_0^x dy\exp\left(\epsilon y-2\int_{x-y}^x  \tilde{\eta}_{t} \,dt \right)
\notag \\
&\,\times
 \delta\tilde{W}^{(n)}_{x-y}(e^{-y}\tilde{\bm k}_1,\cdots,e^{-y}\tilde{\bm k}_n) .
\label{dtGamma}
\end{align}
Let us take the limit $x\rightarrow \infty$ ($\Lambda\rightarrow 0$) in Eq.\ (\ref{dtGamma}) with noting Eq.\ (\ref{eta-def}), 
and make a change of variables as $y=-\ln\lambda$. 
We thereby obtain the ${\bm k}$-dependent vertices for $x\rightarrow\infty$ as
\begin{align}
\delta \tilde{\Gamma}^{(n)}_{\infty}(\tilde{\bm k}_1,\cdots,\tilde{\bm k}_n)=
\int_0^1 \frac{d\lambda}{ \lambda^{1+\epsilon-2\eta}}
 \delta\tilde{W}^{(n)}_{\infty}(\lambda\tilde{\bm k}_1,\cdots,\lambda\tilde{\bm k}_n).
\label{dtGamma_infty}
\end{align}
Let us substitute Eqs.\ (\ref{g_Lambda-asymp}) and (\ref{tildes}) into Eq.\ (\ref{W^(2)-2}), take the limit $\Lambda\rightarrow 0$ subsequently, and use Eq.\ (\ref{dtGamma_infty}).
The procedure yields analytic expressions for $(\delta \tilde{W}^{(2{\rm a})}_\infty,\delta \tilde{W}^{(2{\rm b})}_\infty,\delta \tilde{W}^{(2{\rm c})}_\infty)$ that are free from the renormalization factors $(\Lambda,z_{\Lambda,-})$ as
\begin{subequations}
\begin{align}
&\,
\delta \tilde{W}_{\infty}^{(2{\rm a})}(\tilde{k})
\notag \\
=&\, -\frac{1}{2\beta\,\Psi}\int_0^1 \frac{d\lambda}{\lambda^{1+\epsilon-2\eta}}\left[2\delta\tilde{W}^{(3)}_{\infty}(\lambda\tilde{\bm k},{\bm 0};-\lambda\tilde{\bm k})\right.
\notag \\
&\,
\left.-\delta\tilde{W}^{(3)}_{\infty}(\lambda\tilde{\bm k},-\lambda\tilde{\bm k},{\bm 0}) \right]\int\frac{d^d\tilde{q}}{(2\pi)^d}\left[ 2\dot{\tilde{G}} (\tilde{q})-\dot{\tilde{F}} (\tilde{q})\right],
\label{tW^(2a)}
\end{align}
\begin{align}
&\,\delta\tilde{W}_{\infty}^{(2{\rm b})}(\tilde{k})
\notag \\
=&\,\frac{1}{2\beta}\int\frac{d\lambda}{\lambda^{1+\epsilon-2\eta}}
\int\frac{d^d\tilde{q}}{(2\pi)^d}
\Bigl[2\delta\tilde{W}^{(4)}_{\infty}(\lambda\tilde{\bm k},\lambda\tilde{\bm q};-\lambda\tilde{\bm k},-\lambda\tilde{\bm q}) 
\notag \\
&\,
\times\dot{\tilde{G}}(\tilde{q})
+\delta \tilde{W}^{(4)}_{\infty}(\lambda\tilde{\bm k},-\lambda\tilde{\bm k};\lambda\tilde{\bm q},-\lambda\tilde{\bm q})\dot{\tilde{F}}(\tilde{q}) \Bigr] ,
\label{tW^(2b)}
\end{align}
\begin{align}
&\,\delta \tilde{W}^{(2{\rm c})}_\infty(\tilde{k})
\notag \\
=&\,2g_*\int\frac{d^d \tilde{q}}{(2\pi)^dK_d}
\Bigl[\tilde{F}(|\tilde{\bm k}+\tilde{\bm q}|)\delta(\tilde{q}-1)-\varTheta(|\tilde{\bm k}+\tilde{\bm q}|-1)
\notag \\
&\,\times \dot{\tilde{F}}(\tilde{q})\Bigr]
\notag \\
&\, +\frac{2}{\beta \Psi}\int\frac{d\lambda}{\lambda^{1+\epsilon-2\eta}}\int\frac{d^d \tilde{q}}{(2\pi)^d}
\Bigl\{\delta \tilde{W}^{(3)}_{\infty}(\lambda\tilde{\bm q},-\lambda\tilde{\bm k}-\lambda\tilde{\bm q};\lambda\tilde{\bm k})
\notag \\
&\,\times
\left[\tilde{F}(|\tilde{\bm k}+\tilde{\bm q}|)\delta(\tilde{q}-1)-\varTheta(|\tilde{\bm k}+\tilde{\bm q}|-1)\dot{\tilde{F}}(\tilde{q})\right]
\notag \\
&\, + \Bigl[\delta \tilde{W}^{(3)}_{\infty}(\lambda\tilde{\bm k},\lambda\tilde{\bm q};-\lambda\tilde{\bm k}-\lambda\tilde{\bm q})
\notag \\
&\,
-\delta \tilde{W}^{(3)}_{\infty}(\lambda\tilde{\bm k},-\lambda\tilde{\bm k}-\lambda\tilde{\bm q};\lambda\tilde{\bm q})
\Bigr]\Bigl[\varTheta(|\tilde{\bm k}+\tilde{\bm q}|-1)\dot{\tilde{F}}(\tilde{q})
\notag \\
&\,
+\tilde{F}(|\tilde{\bm k}+\tilde{\bm q}|)\delta(\tilde{q}-1)\Bigr]\Bigr\} ,
\label{tW^(2c)}
\end{align}
\end{subequations}
with
\begin{subequations}
\begin{align}
\Psi\equiv\lim_{\Lambda\rightarrow 0}\Psi_\Lambda,
\end{align}
\begin{align}
\tilde{\dot{G}}(\tilde{q})=\tilde{\dot{F}}(\tilde{q})=\frac{\delta(\tilde{q}-1)}{2},\hspace{5mm}
\tilde{F}(\tilde{q})=-\frac{\varTheta(\tilde{q}-1)}{2\tilde{q}^2}.
\label{tGtF}
\end{align}
\end{subequations}

\subsection{The 2b-4d contribution to $\eta$}
\label{Sec5.4}

\begin{figure}[t]
\begin{center}
\includegraphics[width=0.95\linewidth]{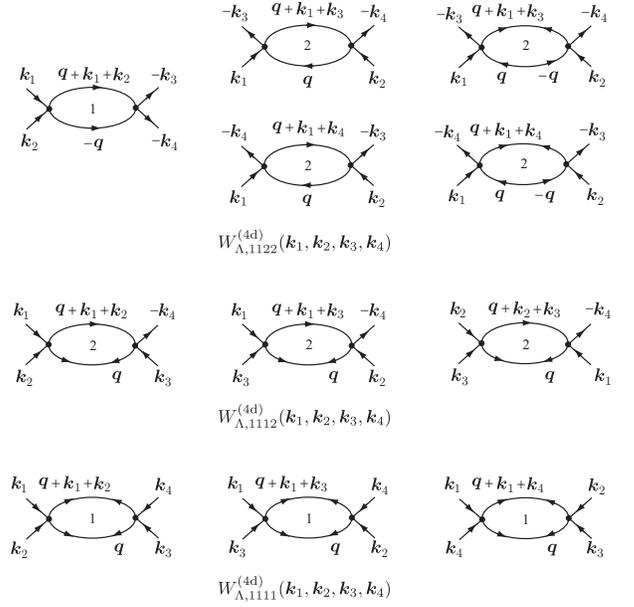}
\end{center}
\caption{Feynman diagrams for $W^{(4{\rm d})}_{\Lambda,j_1\cdots j_n}({\bm k}_1,{\bm k}_2,{\bm k}_3,{\bm k}_4)$ 
representing the fourth term of Eq.\ (\ref{W^(4)-def}) within the approximation of Eq.\ (\ref{Gamma^(n)-approx}). The dash for distinguishing $\hat{\dot{G}}_\Lambda$ from $\hat{G}_\Lambda$ is omitted for simplicity.
The numbers inside the closed internal lines denote the relative weights.
\label{Fig4}}
\end{figure}

We here focus on the ${\bm k}$ dependence of the vertex in Fig.\ \ref{Fig1} (2b)
that originates from the process of Fig.\ \ref{Fig1} (4d) describing the fourth term of Eq.\ (\ref{W^(4)-def});
we denote the corresponding contribution to Eqs.\ (\ref{tW^(2b)}) and (\ref{eta-def2}) as $\delta\tilde{W}^{(2{\rm b}4{\rm d})}_{\infty}$ and $\eta^{(2{\rm b}4{\rm d})}$,
respectively.
With the approximation of Eq.\ (\ref{Gamma^(n0)}) for the vertices, the 4d contribution to Eq.\ (\ref{deltaW^(4)})
is expressible diagrammatically as Fig.\ \ref{Fig4}.
Retaining only the $\omega_\ell=0$ branch
and approximating $G_{\Lambda,12}(k)\approx -G_{\Lambda,11}(k)\equiv F_\Lambda(k)$ based on Eqs.\ (\ref{hatGdG_Lambda}) and (\ref{calG2}),
we obtain
\begin{subequations}
\label{W^(4d)'s}
\begin{align}
&\, W^{(4{\rm d})}_{\Lambda,1122}({\bm k}_1,{\bm k}_2,{\bm k}_3,{\bm k}_4)
\notag \\
=&\,  (2g_\Lambda)^2 \bigl[\chi_{\Lambda,F\dot{F}}(|{\bm k}_1+{\bm k}_2|)+4\chi_{\Lambda,F\dot{F}}(|{\bm k}_1+{\bm k}_3|)
\notag \\
&\,+4\chi_{\Lambda,F\dot{F}}(|{\bm k}_1+{\bm k}_4|)
\bigr] \,\delta_{{\bm k}_1+{\bm k}_2+{\bm k}_3+{\bm k}_4,{\bm 0}},
\label{W^(4d)'s-1}
\end{align}
\begin{align}
&\, W^{(4{\rm d})}_{\Lambda,1112}({\bm k}_1,{\bm k}_2,{\bm k}_3,{\bm k}_4)
\notag \\
=&\,  -2(2g_\Lambda)^2 \bigl[\chi_{\Lambda,F\dot{F}}(|{\bm k}_1+{\bm k}_2|)+\chi_{\Lambda,F\dot{F}}(|{\bm k}_1+{\bm k}_3|)
\notag \\
&\,+\chi_{\Lambda,F\dot{F}}(|{\bm k}_1+{\bm k}_4|)
\bigr] \,\delta_{{\bm k}_1+{\bm k}_2+{\bm k}_3+{\bm k}_4,{\bm 0}},
\end{align}
\begin{align}
&\, W^{(4{\rm d})}_{\Lambda,1111}({\bm k}_1,{\bm k}_2,{\bm k}_3,{\bm k}_4)
\notag \\
=&\,  (2g_\Lambda)^2 \bigl[\chi_{\Lambda,F\dot{F}}(|{\bm k}_1+{\bm k}_2|)+\chi_{\Lambda,F\dot{F}}(|{\bm k}_1+{\bm k}_3|)
\notag \\
&\,+\chi_{\Lambda,F\dot{F}}(|{\bm k}_1+{\bm k}_4|)
\bigr] \,\delta_{{\bm k}_1+{\bm k}_2+{\bm k}_3+{\bm k}_4,{\bm 0}},
\end{align}
\end{subequations}
with
\begin{align}
\chi_{\Lambda,F\dot{F}}(k)\equiv\frac{1}{V\beta}\sum_{{\bm q}_1}F_\Lambda(|{\bm k}+{\bm q}_1|)\dot{F}_\Lambda(q_1).
\label{chi_FF}
\end{align}
Next, we substitute Eq.\ (\ref{W^(4d)'s}) into Eq.\ (\ref{deltaW^(4)}), use Eq.\ (\ref{g_Lambda-asymp}),
perform the transformation of Eq.\ (\ref{tildeW^(n)-def}), and subtract the zero-momenta contribution based on Eq.\ (\ref{dW^(3,4)(0)=0}).
We then find that $\delta\tilde{W}^{(4)}_x$ is expressible without $(\Lambda,z_{\Lambda,-})$ as
\begin{align}
&\,\delta\tilde{W}^{(4{\rm d})}_x(\tilde{\bm k}_1,\tilde{\bm k}_2;\tilde{\bm k}_3,\tilde{\bm k}_4)
\notag \\
=&\, g_*^2\frac{4\beta}{K_d}\left[4\tilde{\chi}_{F\dot{F}}(|\tilde{\bm k}_1+\tilde{\bm k}_2|)+7\tilde{\chi}_{F\dot{F}}(|\tilde{\bm k}_1+\tilde{\bm k}_3|)
\right.
\notag \\
&\,\left. +7\tilde{\chi}_{F\dot{F}}(|\tilde{\bm k}_1+\tilde{\bm k}_4|)-18\tilde{\chi}_{F\dot{F}}(0)\right]\delta_{\tilde{\bm k}_1+\tilde{\bm k}_2+\tilde{\bm k}_3+\tilde{\bm k}_4,{\bm 0}},
\label{tW^(4d)}
\end{align}
where $\tilde{\chi}_{F\dot{F}}(\tilde{k})$ is given in terms of $\tilde{F}\equiv -\tilde{G}_{11}$ and $\dot{\tilde{F}}\equiv -\dot{\tilde{G}}_{11}$ in Eq.\ (\ref{tG(k->0)-T>0}) by
\begin{align}
&\,
\tilde{\chi}_{F\dot{F}}(\tilde{k})
\notag \\
=&\,-\frac{1}{4}\int\frac{d^d\tilde{q}_1}{(2\pi)^d K_d} \delta(\tilde{q}_1-1)\frac{\varTheta(|\tilde{\bm k}+\tilde{\bm q}_1|-1)}{|\tilde{\bm k}+\tilde{\bm q}_1|^2}
\notag \\
=&\, -\frac{\varGamma\left(\frac{d}{2}\right)}{4\pi^{1/2}\varGamma\left(\frac{d-1}{2}\right)}\int_0^{\arccos(-\tilde{k}/2)}d\theta_1\frac{\sin^{d-2}\theta_1}{1+\tilde{k}^2+2\tilde{k}\cos\theta_1} .
\label{tchi}
\end{align}
The second expression has been obtained by expressing $|\tilde{\bm k}+\tilde{\bm q}_1|=(1+\tilde{k}^2+2\tilde{k}\cos\theta_1)^{1/2}$ for $\tilde{q}_1=1$
in terms of the angle $\theta_1$ between ${\bm q}_1$ and ${\bm k}$,
and noting $\varTheta(|\tilde{\bm k}+\tilde{\bm q}_1|-1)=\varTheta(\cos\theta_1+\tilde{k}/2)$.
The normalization constant $\varGamma(d/2)/\pi^{1/2}\varGamma((d-1)/2)$ originates from integrating the Jacobian factor
$\sin^{d-2}\theta_1$ of the spherical coordinates over $\theta_1\in[0,\pi]$.

Let us substitute Eq.\ (\ref{tW^(4d)}) into Eq.\ (\ref{tW^(2b)}) and perform the differentiation of Eq.\ (\ref{eta-def2}). 
We thereby obtain the 2b-4d contribution to the exponent as
\begin{align}
\eta^{(2{\rm b}4{\rm d})}=&\, 18 g_*^2\int_0^1 \frac{d\lambda}{\lambda^{1+\epsilon-2\eta}} \int\frac{d^d\tilde{q}}{(2\pi)^dK_d}\delta(\tilde{q}-1)
\notag \\
&\,\times  \left.  \frac{\partial^2
\tilde{\chi}_{F\dot{F}}(\lambda|\tilde{\bm k}+\tilde{\bm q}|)}{\partial \tilde{k}^2}\right|_{\tilde{k}=0},
\label{eta^(2b4d)-0}
\end{align}
The second derivative in Eq.\ (\ref{eta^(2b4d)-0}) can be calculated
by expressing $|\tilde{\bm k}+\tilde{\bm q}|=(1+\tilde{k}^2+2\tilde{k}\cos\theta_q)^{1/2}$ for $\tilde{q}=1$ and 
using the chain rule as
\begin{align}
&\,\left.\delta(\tilde{q}-1)\frac{\partial^2\tilde{\chi}_{F\dot{F}}(\lambda|\tilde{\bm k}+\tilde{\bm q}|)}{\partial\tilde{k}^2}\right|_{\tilde{k}=0}
\notag \\
=&\, \delta(\tilde{q}-1)\left[\tilde{\chi}_{F\dot{F}}'(\lambda)\lambda(1-\cos^2\theta_q)+ \tilde{\chi}_{F\dot{F}}''(\lambda)\lambda^2 \cos^2\theta_q\right] .
\label{chi''(0)}
\end{align}
Let us substitute Eq.\ (\ref{chi''(0)}) into Eq.\ (\ref{eta^(2b4d)-0}) and integrate over $\lambda$ and over the unit sphere in $d$ dimensions
by using $\langle\cos^2\theta_q\rangle_{\tilde{q}=1}=1/d$, setting $d=4$, and approximating $\lambda^{-\epsilon+2\eta}\approx 1$
to the leading order in $\epsilon$.
We thereby obtain
\begin{align}
\eta^{(2{\rm b}4{\rm d})}=&\,\frac{9}{2}g_*^2 \int_0^1 d\lambda \left[3 \tilde{\chi}_{F\dot{F}}'(\lambda)+\tilde{\chi}_{F\dot{F}}''(\lambda)\lambda\right]= \frac{3}{2}\epsilon^2 ,
\label{eta^(2b4d)}
\end{align}
where we have used $2[\tilde{\chi}_{F\dot{F}}(1)-\tilde{\chi}_{F\dot{F}}(0)]+\tilde{\chi}_{F\dot{F}}'(1)=1/12$ for $d=4$, as can be shown 
elementarily based on Eq.\ (\ref{tchi}), and also Eq.\ (\ref{g_Lambda-asymp}).

Equation (\ref{eta^(2b4d)}) indicates emergence of the exponent $\eta>0$ below $d_{\rm c}=4$ dimensions
given for $d\lesssim 4$ as $\eta\propto \epsilon^2$  in terms of $\epsilon\equiv 4-d$,
in exactly the same way as for the case of the O(2) symmetric $\phi^4$ model at the second-order transition point outlined in Appendix\ref{App4}.

\subsection{The 2c-0 contribution to $\eta$}
\label{subsec:5.5}

We here show that the first term on the right-hand side of Eq.\ (\ref{tW^(2c)}), which 
is apparently proportional to $g_*=2\epsilon$, also gives a contribution of  O($\epsilon^2$) to $\eta$.
Substituting Eq.\ (\ref{tGtF}), we can transform the integral in a way similar to Eq.\ (\ref{tchi}) as
\begin{align}
&\,\delta \tilde{W}^{(2{\rm c}0)}_\infty(\tilde{k})
\notag \\
\equiv &\,-
g_*\int\frac{d^d \tilde{q}}{(2\pi)^dK_d}\delta(\tilde{q}-1)\varTheta(|\tilde{\bm k}+\tilde{\bm q}|-1)
\Biggl(\!\frac{1}{|\tilde{\bm k}+\tilde{\bm q}|^2}+1\!\Biggr)
\notag \\
= &\,- g_*\frac{\varGamma\left(\frac{d}{2}\right)}{\pi^{1/2}\varGamma\left(\frac{d-1}{2}\right)}\int_{-\tilde{k}/2}^1 dt_q
\left(\frac{1}{1+\tilde{k}^2+2\tilde{k}t_q}+1\right) 
\notag \\
&\,\times (1-t_q^2)^{(d-3)/2} .
\end{align}
where $t_q\equiv \cos\theta_q$ with $\theta_q$ denoting the angle between $\tilde{\bm k}$ and $\tilde{\bm q}$.
The corresponding contribution to Eq.\ (\ref{eta-def2}) is calculated as
\begin{align}
&\,\eta^{(2{\rm c}0)}
\notag \\
\equiv &\,\left.\frac{1}{2}\frac{\partial^2\delta \tilde{W}^{(2{\rm c}0)}_\infty(\tilde{k})}{\partial\tilde{k}^2}\right|_{\tilde{k}=0}
\notag \\
=&\, -g_*\frac{\varGamma\left(\frac{d}{2}\right)}{\pi^{1/2}\varGamma\left(\frac{d-1}{2}\right)}\int_{0}^{\pi/2} d\theta_q \sin^{d-2}\theta_q(-1+4\cos^2\theta_q)
\notag \\
=&\,  -g_*\frac{4-d}{2d}\approx -\frac{\epsilon^2}{4} .
\label{eta^(2c0)}
\end{align}
where we have used Eq.\ (\ref{g_Lambda-asymp}).

\section{Comparisons with Previous Studies}
\label{Sec6}

We here compare the present (i) formulation and (ii) results with those of the relevant previous ones. 

(i) Equation (\ref{Gamma^(n)-eq}) with Fig.\ \ref{Fig1} looks almost identical with Eqs.\ (20)-(22) and Fig.\ 1 of Sch\"utz and Kopietz.\cite{SK06}
The distinct features are summarized as follows. 
First, Eq.\ (\ref{Gamma^(n)-eq}) is designed to be solved in such a way as to be compatible with Eq.\ (\ref{GoldstoneI})
by adopting the parametrization of Eq.\ (\ref{Gamma^(n)-parametrize}), for example.
Indeed, incorporating the identities of Eq.\ (\ref{GoldstoneI}) into the equations results in a 
substantial reduction of the flow parameters and also simplification of the equations to be solved, 
which also obeys Goldstone's theorem (I) automatically,
as exemplified in Sec.\ 4.
Second, since we adopt a simple Legendre transformation in Eq.\ (\ref{Gamma-def}),
the extremal condition $\Gamma_{\Lambda,j}^{(1)}(\kappa)=0$ is manifest in our formulation.
On the other hand, it is imposed by Sch\"utz and Kopietz together with the additional condition of Eq.\ (26) in their paper.\cite{SK06}
Hence, our equations are easier to handle in practical calculations.

(ii) Solving Eq.\ (\ref{Gamma^(n)-eq}), we have shown that that the anomalous self-energy $\Delta(0)$ at zero energy-momenta vanishes
for $d\leq 4$ ($T>0$) and $d\leq 3$ ($T=0$)  dimensions as Eqs.\ (\ref{g_Lambda-asymp}) and (\ref{g_Lambda-asymp2}), respectively.
The latter result, i.e., $\Delta(0)=0$ at $T=0$ for  $d\leq 3$, is simply a reproduction of the Nepomnyashchi\u{i} identity,\cite{SHK09}
which also has been reproduced by Popov and Seredniakov,\cite{Popov79} 
Castellani {\em et al}.,\ \cite{CCPS97,PCCS04} Dupuis and Sengupta,\cite{DS07}
and Sinner {\em et al}.\cite{SHK09}
The fact indicates that the infrared divergences that emerge in the perturbative treatment of Bose-Einstein condensates 
are not an artifact but result in the substantial modification of the prediction by the weak-coupling Bogoliubov theory.\cite{Bogoliubov47}
The present paper has shown how the Nepomnyashchi\u{i} identity can be extended to finite temperatures.
The latter three studies,\cite{CCPS97,DS07,SHK09}  which are also based on the renormalization group, 
also discuss persistence of a sound wave in the single-particle channel at $T=0$,
which is distinct from the Bogoliubov mode,\cite{CCPS97}
by incorporating second-order frequency renormalization factors such as ${\partial^2\Sigma(k,\omega)}/{\partial\omega^2}\bigl|_{k,\omega=0}$
besides the standard ones in the first order, e.g., ${\partial\Sigma(k,\omega)}/{\partial\omega}\bigl|_{k,\omega=0}$ and  
${\partial\Sigma(k,\omega)}/{\partial k}\bigl|_{k,\omega=0}$.
It should be noted, however, that their arguments cannot be used at finite temperatures where the Matsubara frequencies
become discrete. Moreover, they incorporate only the contribution of Fig.\ 1 (4d) with two four-point vertices
into their flow equations, while those of Fig.\ 1 (4e) and (4f) with three-point vertices are equally important
as revealed here. Hence, their statement is yet to be confirmed, especially at finite temperatures.

On the other hand, our analysis in Sec.\ 4 focuses on the main flow parameter $g_\Lambda$ alone 
and discusses how it vanishes for $\Lambda\rightarrow 0$ just below the critical dimensions $d_{\rm c}=4$ ($T>0$) and 
$d_{\rm c}=3$ ($T=0$) based on Eq.\ (\ref{Gamma^(n)-eq}) for $n=2$, where
all the diagrams for $W_\Lambda^{(2)}$ in Fig.\ \ref{Fig1} have been incorporated without any omission.
Moreover, the analysis in Sec.\ 4
can be regarded as a perturbative treatment on the anomalous dimension $\eta$ of the correlation function at finite temperatures by
assuming the smallness $\eta\propto \epsilon^2$ as compared with $g_*\propto \epsilon$ for $(\epsilon\equiv 4-d \ll 1$).
Indeed, it is shown in Appendix \ref{App4} that the strategy can reproduce the anomalous dimension of the correlation function for
the O(2) symmetric $\phi^4$ model correctly near its critical dimension.
The analysis on $\eta$ with first two terms suggests that $\eta$ emerges as $\eta\propto \epsilon^2$.
The result of a complete calculation of $\eta$ for $d\lesssim 4$ at finite temperatures will be reported in a separate paper shortly.

\section{Summary}
\label{Sec7}

We have derived a set of exact renormalization-group equations for interacting 
Bose-Einstein condensates as Eq.\ (\ref{Gamma^(n)-eq}) with Eq.\ (\ref{W^(n)-def}).
As shown in Sec.\ \ref{subsec:3.3}, they automatically obey Goldstone's theorem (I) when the initial vertices satisfy Eq.\ (\ref{GoldstoneI}),
such as Eq.\ (\ref{Gamma^(nB)}) by the Bogoliubov approximation.
Given in terms of a fixed chemical potential $\mu$, the strength of the interaction in the present formalism can be monitored
by the dimensionless ratio $\mu/k_{\rm B}T_0$, where $T_0$ is the transition temperature of the non-interacting Bose gas in $d$ dimensions.
They will form a new basis for detailed analytical and numerical studies on Bose-Einstein condensates,
especially those at low energies where the perturbation theory has encountered difficulties of infrared divergences\cite{GN64}
or emergence of an unphysical energy gap.\cite{HM65}

Using them, it has been shown that (i) the interaction $g_\Lambda$ vanishes 
in the infrared limit $\Lambda\rightarrow 0$ below $d_{\rm c}=4$ ($d_{\rm c}=3$)
dimensions at finite temperatures (zero temperature) in Sec.\ \ref{subsec4.3} (Sec.\ \ref{subsec4.4}).
This vanishing of $g_\Lambda$ is also accompanied by the development of the exponent $\eta>0$,
as shown in Sec.\ \ref{Sec5}, which for $d\lesssim d_{\rm c}$ is proportional to $\epsilon^2\equiv(4-d)^2$.
A complete calculation of the prefactor will be reported in a separate paper.

\section*{Acknowledgment}
This work is supported by Yamada Science Foundation.

\appendix

\section{One-particle Green's functions}
\label{App1}

Green's functions $G^{(2)}_{j_1j_2}(\xi_1,\xi_2)$ in the $\xi=({\bm r},\tau)$ space
generally satisfy\cite{Kita-Text}
\begin{align}
\left[G^{(2)}_{j_1j_2}(\xi_1,\xi_2)\right]^*=G^{(2)}_{3-j_2,3-j_1}({\bm r}_2\tau_1,{\bm r}_1\tau_2) .
\label{G^*-symm}
\end{align}
Upon the Fourier transform
\begin{align*}
&\,
G^{(2)}_{j_1j_2}(\xi_1,\xi_2)
\notag \\
=&\,\frac{1}{V \beta}\sum_{\kappa_1\kappa_2} G^{(2)}_{j_1j_2}(\kappa_1,\kappa_2)
e^{i{\bm k}_1\cdot {\bm r}_1-i\omega_{\ell_1}\tau_1+i{\bm k}_2\cdot {\bm r}_2-i\omega_{\ell_2}\tau_2},
\end{align*}
Eq.\ (\ref{G^*-symm}) translates into
\begin{subequations}
\label{G(k1,k2)-symm}
\begin{align}
\left[G^{(2)}_{j_1j_2}(-\kappa_1,-\kappa_2)\right]^*=G^{(2)}_{3-j_2,3-j_1}({\bm k}_2\omega_{\ell_1},{\bm k}_1\omega_{\ell_2}) .
\end{align}
It also follows from Eq.\ (\ref{G^(n)}) that 
\begin{align}
\hat{G}^{(2)}_{j_1j_2}(\kappa_1,\kappa_2)=\hat{G}^{(2)}_{j_2j_1}(\kappa_2,\kappa_1) 
\end{align}
\end{subequations}
holds.
Substituting Eq.\ (\ref{G^(2)-G}) into Eq.\ (\ref{G(k1,k2)-symm}), we find that $G_{j_1j_2}(\kappa)$ satisfy
$G_{j_1j_2}^*({\bm k},i\omega_\ell)=G_{3-j_2,3-j_1}({\bm k},-i\omega_\ell)$ and
$G_{j_1j_2}(\kappa)=G_{j_2j_1}(-\kappa)$.
Moreover, since the system is isotropic here, 
Green's functions depend on ${\bm k}$ only through $k\equiv|{\bm k}|$.
Hence, the two symmetry relations can be simplified further into
\begin{subequations}
\label{G_ij-symm}
\begin{align}
G_{j_1j_2}^*(k,i\omega_\ell)=&\, G_{3-j_2,3-j_1}(k,-i\omega_\ell) ,
\label{G_ij-symm1}
\\
G_{j_1j_2}(k,i\omega_\ell)=&\,G_{j_2j_1}(k,-i\omega_\ell).
\label{G_ij-symm2}
\end{align} 
\end{subequations}
It follows from Eq.\ (\ref{G_ij-symm2}) that the diagonal elements are even functions of $i \omega_\ell$, 
which implies that they are real functions
in the gauge where $\Psi$ is real.
Combining the results with Eq.\ (\ref{G_ij-symm1}), we obtain the symmetry relations
\begin{subequations}
\begin{align}
G_{11}(k,i\omega_\ell)=&\,G_{11}(k,-i\omega_\ell)=G_{11}^*(k,i\omega_\ell)=G_{22}(k,i\omega_\ell) 
\end{align}
for the diagonal elements. We can also conclude that
\begin{align}
G_{12}(k,i\omega_\ell)=G_{21}(k,-i\omega_\ell)=G_{21}^*(k,i\omega_\ell) 
\end{align} 
\end{subequations}
holds for the off-diagonal elements. Writing $G_{11}(k,i\omega_\ell)=-F(k,i\omega_\ell)$
and $G_{12}(k,i\omega_\ell)=G(k,i\omega_\ell)$, we arrive at Eq.\ (\ref{hatG}).

\section{Derivation of Eq.\ (\ref{dlnZ})}
\label{App2}

Let us express the non-interacting part of Eq.\ (\ref{S-k}) with a cutoff $\Lambda$ in a matrix form as
\begin{align}
S_{0\Lambda}=&\,-\beta \sum_{\kappa}\psi_2 (-\kappa)G_{0\Lambda}^{-1}(\kappa)\psi_1 (\kappa)
\notag \\
=&\, -\frac{\beta}{2}\sum_{j_1,\kappa_1}\sum_{j_2,\kappa_2}\psi_{j_1}(\kappa_1) \left(\hat{G}_{0\Lambda}^{(2)\,-1}\right)_{j_1\kappa_1,j_2\kappa_2}\psi_{j_2}(\kappa_2) ,
\end{align}
where $\hat{G}_{0\Lambda}^{(2)\,-1}$ is a matrix defined by 
\begin{align}
\left(\hat{G}_{0\Lambda}^{(2)\,-1}\right)_{j_1\kappa_1,j_2\kappa_2}=\delta_{\kappa_1+\kappa_2,0}
\begin{bmatrix}
0 & G_{0\Lambda}^{-1}(-\kappa_2) \\
G_{0\Lambda}^{-1}(\kappa_2) & 0 
\end{bmatrix}_{j_1j_2} .
\label{hatG_0^(2)-1}
\end{align}
Subsequently, we transform $\partial_\Lambda \ln Z_\Lambda[J]$ as
\begin{align}
&\,\partial_\Lambda \ln Z_\Lambda[J]
\notag \\
=&\, \frac{1}{Z_\Lambda[J]} \int D[\psi] \, 
\frac{\beta}{2} \sum_{j_1,\kappa_1}\sum_{j_2,\kappa_2} \psi_{j_1} (\kappa_1) 
\left(\partial_\Lambda \hat{G}_{0\Lambda}^{(2)\,-1}\right)_{j_1\kappa_1,j_2\kappa_2}
\notag \\
&\,\times \psi_{j_2} (\kappa_2)\exp\left[-S_{\Lambda}+\sum_{j,\kappa} J_j(\kappa) \psi_j(\kappa) \right] 
\notag \\
=&\, \frac{1}{2} \sum_{j_1,\kappa_1}\sum_{j_2,\kappa_2}\left(\partial_\Lambda \hat{G}_{0\Lambda}^{(2)\,-1}\right)_{j_1\kappa_1,j_2\kappa_2}
\beta \frac{\delta^2\ln Z_\Lambda[J]}{\delta J_{j_2}(\kappa_2)\delta J_{j_1}(\kappa_1)}
\notag \\
=&\,- \frac{1}{2} {\rm Tr} \,\left(\partial_\Lambda \hat{G}_{0\Lambda}^{(2)\,-1}\right)
\hat{G}_\Lambda^{(2)}\sum_{\nu=0}^\infty \left(\hat{U}_\Lambda\hat{G}_\Lambda^{(2)}\right)^\nu,
\label{dlnZ_L}
\end{align}
where Tr denotes trace, and we have substituted Eq.\ (\ref{dlnZ-U}).
Let us insert the unit matrix $\hat{G}^{(2)\,-1}\hat{G}^{(2)}$ immediately after Tr in Eq.\ (\ref{dlnZ_L}), separate the $\nu=0$ term from the sum, and 
and make a change of variables $\nu=\nu'+1$ in the remaining sum over $\nu=1,2,\cdots$.
We thereby obtain Eq.\ (\ref{dlnZ}) in terms of the quantity
\begin{align}
\hat{\dot{G}}_\Lambda^{(2)} \equiv -\hat{G}_\Lambda^{(2)}\left(\partial_\Lambda \hat{G}_{0\Lambda}^{(2)\,-1}\right)\hat{G}_\Lambda^{(2)}.
\label{dotG-A}
\end{align}
The remaining task is to transform Eq.\ (\ref{dotG-A}) into Eq.\ (\ref{dotG}).
To this end,  we differentiate the identity $\hat{G}_{0\Lambda}^{(2)}\hat{G}_{0\Lambda}^{(2)\,-1}=\hat{1}$ with respect to $\Lambda$;
the result is expressible as
$\partial_\Lambda \hat{G}_{0\Lambda}^{(2)\,-1}=-\hat{G}_{0\Lambda}^{(2)\,-1}\left(\partial_\Lambda \hat{G}_{0\Lambda}^{(2)}\right)\hat{G}_{0\Lambda}^{(2)\,-1}$. Substituting it into Eq.\ (\ref{dotG-A}) and writing $\hat{G}_\Lambda^{(2)}\hat{G}_{0\Lambda}^{(2)\,-1}=
\left(\hat{G}_{0\Lambda}^{(2)\,-1}-\hat{\Sigma}_{\Lambda}^{(2)}\right)^{-1}\hat{G}_{0\Lambda}^{(2)\,-1}=
\left(\hat{1}-\hat{G}_{0\Lambda}^{(2)}\hat{\Sigma}_{\Lambda}^{(2)}\right)^{-1}$
based on Eq.\ (\ref{Sigma^(2)}), we arrive at Eq.\ (\ref{dotG}).

\section{Proof of Eq.\ (\ref{W^(2)-identity})}
\label{App3}

We prove Eq.\ (\ref{W^(2)-identity}) as $C_2=0$ in terms of
\begin{align}
C_2\equiv &\,\lim_{\kappa\rightarrow 0} \,\biggl[W_{\Lambda,12}^{(2)}(\kappa,-\kappa)-W_{\Lambda,11}^{(2)}(\kappa,-\kappa)\biggr]
\notag \\
\equiv &\, C_{2{\rm a}}+C_{2{\rm b}}+C_{2{\rm c}1}+C_{2{\rm c}2} ,
\label{C}
\end{align}
where the four terms are defined based on Eq.\ (\ref{W^(2)-def}) by
\begin{subequations}
\begin{align}
C_{2{\rm a}}\equiv &\,-\sum_{jj_1}\,(-1)^{j}\,\Gamma_{\Lambda,1jj_1}^{(3)}(0,0,0)
\partial_\Lambda\Psi_{\Lambda},
\label{C_1}
\\
C_{2{\rm b}}\equiv&\,\frac{1}{2V \beta}{\rm Tr} \,\sum_{j}
(-1)^{j}\,\hat\Gamma^{(4)}_{\Lambda,1j}(0,0)\,\hat{\dot{G}}_\Lambda^{(2)},
\label{C_2}
\\
C_{2{\rm c}1}\equiv&\,\frac{1}{2V \beta}{\rm Tr}\lim_{\kappa\rightarrow 0}
\hat\Gamma^{(3)}_{\Lambda,1}(\kappa)\,\hat{G}_\Lambda^{(2)}\,\sum_{j}(-1)^j 
\,\hat\Gamma^{(3)}_{\Lambda,j}(-\kappa)\,
\hat{\dot{G}}_\Lambda^{(2)},
\label{C_3}
\\
C_{2{\rm c}2}\equiv&\,\frac{1}{2V \beta}{\rm Tr} \lim_{\kappa\rightarrow 0}\,\sum_{j}(-1)^j 
\,\hat\Gamma^{(3)}_{\Lambda,j}(-\kappa)\,\hat{G}_\Lambda^{(2)}\,
\hat\Gamma^{(3)}_{\Lambda,1}(\kappa)\,
\hat{\dot{G}}_\Lambda^{(2)}.
\label{C_4}
\end{align}
\end{subequations}
First, we transform $C_{2{\rm a}}$ using Eqs.\ (\ref{HP}) and (\ref{dPsi-eq}) as
\begin{align}
C_{2{\rm a}}=&\, -\frac{2}{\Psi_\Lambda}\Gamma^{(2)}_{\Lambda,11}(0,0)\,\partial_\Lambda\Psi_{\Lambda}
\notag \\
=&\, -\frac{1}{2V \beta \,\Psi_\Lambda}\sum_{\kappa}\sum_{jj'}\Gamma^{(3)}_{\Lambda,1j'j}(0,-\kappa,\kappa)\,\dot{G}_{\Lambda,jj'}(\kappa) ,
\label{C_1-1}
\end{align}
where we have also expressed $\hat{\dot{G}}^{(2)}$ as Eq.\ (\ref{dotG^(2)-dotG}).
Next, we transform Eq.\ (\ref{C_2}) using Eq.\ (\ref{GoldstoneI}) for $n=3$ as
\begin{align}
C_{2{\rm b}} =&\,\frac{1}{2V \beta}\sum_{\kappa_1}\sum_{j_1j_1'} \sum_j\,(-1)^j\,
\Gamma^{(4)}_{\Lambda,1j j_1'j_1}(0,0,-\kappa_1,\kappa_1)
\notag \\
&\,\times \dot{G}_{\Lambda,j_1j_1'}(\kappa_1) 
\notag \\
=&\, \frac{1}{2V \beta\,\Psi_\Lambda}\sum_{\kappa_1}\sum_{j_1j_1'}\,\biggl[1+2\delta_{j_1j_1'}(-1)^{j_1-1}\biggr]\,
\notag \\
&\,\times 
\Gamma^{(3)}_{\Lambda,1j_1'j_1}(0,-\kappa_1,\kappa_1)
\dot{G}_{\Lambda,j_1j_1'}(\kappa_1) ,
\label{C_2-1}
\end{align}
where we have also used $(-1)^{j_1-1}+(-1)^{j_1'-1}=2(-1)^{j_1-1}\delta_{j_1j_1'}$. Third, Eq.\ (\ref{C_3}) is transformed as
\begin{align}
&\,C_{2{\rm c}1}
\notag \\
=&\, \frac{1}{2V \beta} \sum_{\kappa_1}\sum_{j_1j_1'}\sum_{j_2j_2'} \lim_{\kappa\rightarrow 0}
\Gamma^{(3)}_{\Lambda,1j_1'j_1}(\kappa,-\kappa_1-\kappa,\kappa_1)
\notag \\
&\,\times
G_{\Lambda,j_1j_2}(\kappa_1)
\sum_{j=1}^2 (-1)^j\,\Gamma^{(3)}_{\Lambda,j j_2j_2'}(-\kappa,-\kappa_1,\kappa_1+\kappa) 
\notag \\
&\,\times
\dot{G}_{\Lambda,j_2'j_1'}(\kappa_1+\kappa) 
\notag \\
=&\, \frac{1}{V \beta\,\Psi_\Lambda} \sum_{\kappa_1}\sum_{j_1j_1'}\sum_{j_2j_2'} 
\Gamma^{(3)}_{\Lambda,1j_1'j_1}(0,-\kappa_1,\kappa_1)(-1)^{j_2-1} 
\notag \\
&\,\times \Gamma^{(2)}_{\Lambda, j_2j_2}(-\kappa_1,\kappa_1)
\lim_{\kappa\rightarrow 0} G_{\Lambda,j_1j_2}(\kappa_1)\dot{G}_{\Lambda,j_2j_1'}(\kappa_1+\kappa) ,
\label{C_3-0}
\end{align}
where we have safely set $\kappa=0$ in $\Gamma^{(3)}$ and used Eq.\ (\ref{GoldstoneI}) for $n=2$ and 
$(-1)^{j_2-1}+(-1)^{j_2'-1}=2(-1)^{j_2-1}\delta_{j_2j_2'}$.
Substituting Eq.\ (\ref{hatGdG_Lambda}), we can express the above limit of $\kappa\rightarrow 0$ as
\begin{align}
&\,\lim_{\kappa\rightarrow 0} G_{\Lambda,j_1j_2}(\kappa_1)\dot{G}_{\Lambda,j_2j_1'}(\kappa_1+\kappa) 
\notag \\
=&\,-\frac{1}{2}
{\cal G}_{\Lambda,j_1j_2}(\kappa_1){\cal G}_{\Lambda,j_2j_1'}(\kappa_1)\delta(k_1-\Lambda) ,
\label{GG-lim}
\end{align}
where we have taken the limit $k\rightarrow0$ first
and subsequently used $\varTheta(x)\delta(x)=\delta(x)/2$.
Equation (\ref{GG-lim}) is also confirmed to hold by transforming
$\lim_{{\bm k}\rightarrow0}\varTheta(|{\bm k}_1+{\bm k}|-\Lambda)\delta(k_1-\Lambda)
=\lim_{{\bm k}\rightarrow0}\varTheta(\cos\theta_1+k/2\Lambda)\delta(k_1-\Lambda)
=\varTheta(\cos\theta_1)\delta(k_1-\Lambda)$  with $\theta_1$ the angle between ${\bm k}$ and ${\bm k}_1$,
integrating the expression
over $\theta_1\in[0,\pi]$ with the Jacobian factor $\sin^{d-2}\theta_1$, and comparing the result with that obtained without the factor $\varTheta(\cos\theta_1)$.

Now, we can transform Eq.\ (\ref{C_3-0}) further by  using (i) Eq.\ (\ref{GG-lim}), 
(ii) the symmetries ${\cal G}_{\Lambda,11}(\kappa_1)={\cal G}_{\Lambda,22}(\kappa_1)$, ${\cal G}_{\Lambda,21}(\kappa_1)={\cal G}_{\Lambda,12}(-\kappa_1)$, $\Gamma^{(2)}_{\Lambda,22}(-\kappa_1,\kappa_1)=\Gamma^{(2)}_{\Lambda,11}(-\kappa_1,\kappa_1)=\Gamma^{(2)}_{\Lambda,11}(\kappa_1,-\kappa_1)$, and (iii) the identity $\Gamma^{(2)}_{\Lambda,11}(-\kappa_1,\kappa_1)[{\cal G}_{\Lambda,12}(\kappa_1){\cal G}_{\Lambda,21}(\kappa_1)-{\cal G}_{\Lambda,11}(\kappa_1){\cal G}_{\Lambda,22}(\kappa_1)]={\cal G}_{\Lambda,11}(\kappa_1)$
that originates from Eq.\ (\ref{calG}). 
Terms with $j_1'\neq j_1$ are found to cancel out, and we obtain
\begin{align}
C_{2{\rm c}1}=&\,\frac{1}{2V \beta\,\Psi_\Lambda} \sum_{j_1,\kappa_1}(-1)^{j_1}\Gamma_{\Lambda,1j_1j_1}^{(3)}(0,-\kappa_1,\kappa_1)
\notag \\
&\,\times
\dot{G}_{\Lambda,j_1j_1}(\kappa_1).
\label{C_3-1}
\end{align}
Finally, we substitute Eqs.\ (\ref{C_1-1}), (\ref{C_2-1}), (\ref{C_3-1}), and $C_{2{\rm c}2}=C_{2{\rm c}1}$ into Eq.\ (\ref{C}).
We then arrive at $C_2=0$.

We have confirmed that Eq.\ (\ref{W^(3)-identity}) can be proved similarly.

\section{Wilson-Fisher $\epsilon$ expansion}
\label{App4}

\begin{figure}[b]
\begin{center}
\includegraphics[width=0.9\linewidth]{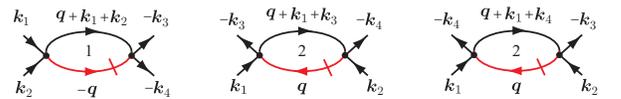}
\end{center}
\caption{Feynman diagrams for $W^{(4)}_{\Lambda,1122}$ of Eq.\ (\ref{W^(4)-def}) at the critical point.
Numbers inside the closed particle lines denote the relative weights.
\label{FigAD}}
\end{figure}

Here, we show how the result of the Wilson-Fisher $\epsilon$ expansion\cite{WF72,Wilson72,Amit,Justin96,KBS10} at the transition point can be reproduced for the present model within the present formalism.

According to the power-counting scheme in the theory of critical phenomena,\cite{Amit,KBS10} 
the only relevant vertex at the transition point is identified to be
\begin{align}
\Gamma_{1122}^{(4)}(0,0,0,0)=\frac{2u_\Lambda}{3} ,
\label{Gamma_1122^(4)}
\end{align}
which is denoted as $2g_\Lambda$ in Eq.\ (\ref{Gamma^(40)}); making a change of variables 
$\psi_1=(\phi_1+i\phi_2)/\sqrt{2}$ and $\bar\psi_2=(\phi_1-i\phi_2)/\sqrt{2}$ with $g_0=u_0/3$ in Eq.\ (\ref{S-def}) yields the standard expression of 
the action for the O(2) symmetric $\phi^4$ model with the bare four-point vertex $u_0$.\cite{Amit,Justin96}
We also take the limit $v_k\rightarrow 0$ in Eq.\ (\ref{G(k->0)}) and set $\omega_\ell=0$ as appropriate at the transition point.
The procedure yields
\begin{align}
G_\Lambda(k)\equiv G_{\Lambda,12}(k)\approx G_{\Lambda,21}(k)=-\frac{z_{\Lambda,12}}{k^2} \varTheta(k-\Lambda),
\label{G-normal}
\end{align}
and $G_{\Lambda,11}(k)=G_{\Lambda,22}(k)=0$.
Let us substitute them into Eq.\ (\ref{W^(4)-def}) for $(j_1,j_2,j_3,j_4)=(1,1,2,2)$,
which is expressible diagrammatically for $\omega_\ell=0$ as Fig.\ \ref{FigAD}.
We then obtain
\begin{align}
&\, W^{(4)}_{\Lambda,1122}({\bm k}_1,{\bm k}_2,{\bm k}_3,{\bm k}_4)
\notag \\
=&\,  \left(\frac{2}{3}u_\Lambda\right)^2 \bigl[\chi_{\Lambda,G\dot{G}}(|{\bm k}_1+{\bm k}_2|)+2\chi_{\Lambda,G\dot{G}}(|{\bm k}_1+{\bm k}_3|)
\notag \\
&\,+2\chi_{\Lambda,G\dot{G}}(|{\bm k}_1+{\bm k}_4|)
\bigr]\,\delta_{{\bm k}_1+{\bm k}_2+{\bm k}_3+{\bm k}_4,{\bm 0}} ,
\label{W^(4d)'s-c}
\end{align}
where $\chi_{\Lambda,G\dot{G}}(k)$ is defined similarly as Eq.\ (\ref{chi_FF}) in terms of Eq.\ (\ref{G-normal}).
Next, we substitute $W_{1122}^{(4)}(0,0,0,0)$ and Eq.\ (\ref{Gamma_1122^(4)}) 
into Eq.\ (\ref{Gamma^(n)-eq}) for $n=4$ at zero momenta.
We thereby obtain the equation for $u_\Lambda$ as
\begin{align}
-\partial_\Lambda u_\Lambda=5\cdot \frac{2}{3} u_\Lambda^2\, \chi_{\Lambda,G\dot{G}}(0) .
\label{dbg-eq}
\end{align}
The quantity $\chi_{\Lambda,G\dot{G}}(0)$ can be calculated in the same way as Eq.\ (\ref{chi(T>0)})
by using Eq.\ (\ref{G-normal}) instead of Eq.\ (\ref{GF-dGF}),
\begin{align}
\chi_{\Lambda,G\dot{G}}(0)=-\frac{z_{\Lambda,12}^2}{2\beta}K_d \Lambda^{-1-\epsilon}.
\end{align}
Let us substitute this expression into Eq.\ (\ref{dbg-eq}) and integrate the resulting equation for 
$d\lesssim 4$. We then find that $u_\Lambda$ behaves for $\Lambda\rightarrow 0$ as
\begin{align}
u_\Lambda = u_* \frac{\beta}{K_d\,z_{\Lambda,12}^2}\Lambda^\epsilon,\hspace{10mm}u_*\equiv \frac{3}{5}\epsilon ,
\label{u_Lambda-sol}
\end{align}
in agreement with the result for the O(2) symmetric $\phi^4$ theory.\cite{Amit,Justin96}
The quantity $u_*$ is called {\it fixed point} in the theory of critical phenomena.\cite{Wilson74,SKMa,Amit,Justin96,KBS10}
Equation (\ref{u_Lambda-sol}), which has been obtained here without the rescaling procedure for the vertices,
clearly indicates that the interaction at zero momenta vanishes at the transition point so as to remove the
infrared divergence.
This fact seems not to have been stated explicitly in the literature.

Next, we focus on the exponent $\eta$ at the critical point.
Let us substitute Eq.\ (\ref{W^(4d)'s-c}) and $W^{(4)}_{\Lambda,1112}=W^{(4)}_{\Lambda,1111}=0$  together with Eq.\ (\ref{u_Lambda-sol})
into Eq.\ (\ref{deltaW^(4)}),
perform the transformation of Eq.\ (\ref{tildes}) with replacing $z_{\Lambda,-}$ by  $z_{\Lambda,12}$,
use the resulting expression of $\delta\tilde{W}^{(4)}_{\infty}(\tilde{\bm k}_1,\tilde{\bm k}_2;\tilde{\bm k}_3,\tilde{\bm k}_4)$ 
and $\dot{\tilde{G}}(\tilde{k}_1)$
in Eq.\ (\ref{tW^(2b)}), and perform the differentiation of Eq.\ (\ref{eta-def2}). 
We thereby obtain the expression for $\eta$ as
\begin{align}
\eta=&\, \frac{2}{3} u_*^2\int_0^1 d\lambda\, \lambda^{-1-\epsilon+2\eta} \int\frac{d^d\tilde{k}_1}{(2\pi)^dK_d}\delta(\tilde{k}_1-1)
\notag \\
&\,\times  \left.  \frac{\partial^2
\tilde{\chi}_{G\dot{G}}(\lambda|\tilde{\bm k}+\tilde{\bm k}_1|)}{\partial \tilde{k}^2}\right|_{\tilde{k}=0} ,
\label{eta^(critical)-0}
\end{align}
where $\tilde{\chi}_{G\dot{G}}(\tilde{k})$ is four times as large as Eq.\ (\ref{tchi}) owing to the difference between 
Eqs.\ (\ref{calG2}) and (\ref{G-normal}).
The two integrals in Eq.\ (\ref{eta^(critical)-0}) can be performed in the same way as Eqs.\ (\ref{tchi})-(\ref{eta^(2b4d)}).
We thereby reproduce the well-known result for the O(2) symmetric $\phi^4$ theory,\cite{Wilson72,Wilson74,Fisher74,Amit,Justin96}
\begin{align}
\eta=&\,\frac{1}{6}u_*^2 \int_0^1 d\lambda \left[3 \tilde{\chi}_{G\dot{G}}'(\lambda)+\tilde{\chi}_{G\dot{G}}''(\lambda)\lambda\right]= \frac{\epsilon^2}{50} .
\label{bareta}
\end{align}

\section{Derivation of Eq.\ (\ref{W^(2c)-2})}
\label{App5}

Using Figs.\ \ref{Fig2} and \ref{Fig3} together with  Eqs.\ (\ref{W^(2)-def}) and (\ref{Gamma^(n)-approx}), 
we obtain $W^{(2{\rm c})}_\Lambda(k)$ as 
\begin{align}
&\,\delta W^{(2{\rm c})}_\Lambda(k)
\notag \\
=&\,\frac{(2g_\Lambda\Psi_\Lambda)^2}{\beta}\int\frac{d^d q}{(2\pi)^d}
\Bigl[G_{\Lambda}(|{\bm k}+{\bm q}|)\dot{G}_{\Lambda}(q)-F_{\Lambda}(|{\bm k}+{\bm q}|)
\notag \\
&\,\times \dot{F}_{\Lambda}(q)\Bigr]+\frac{4g_\Lambda\Psi_\Lambda}{\beta}\int\frac{d^d q}{(2\pi)^d}
\left\{ \delta \Gamma^{(3)}_{\Lambda}({\bm q},-{\bm k}-{\bm q};{\bm k})
\right.
\notag \\
&\,\times
\left[G_{\Lambda}(|{\bm k}+{\bm q}|)\dot{G}_{\Lambda}(q)-F_{\Lambda}(|{\bm k}+{\bm q}|)\dot{F}_{\Lambda}(q)\right]
\notag \\
&\, + \left[\delta \Gamma^{(3)}_{\Lambda}({\bm k},-{\bm k}-{\bm q};{\bm q})
-\delta \Gamma^{(3)}_{\Lambda}({\bm k},{\bm q};-{\bm k}-{\bm q})\right]
\notag \\
&\,\left.\times
\left[G_{\Lambda}(|{\bm k}+{\bm q}|)\dot{F}_{\Lambda}(q)-F_{\Lambda}(|{\bm k}+{\bm q}|)\dot{G}_{\Lambda}(q)\right]\right\} ,
\label{W^(2c)-App}
\end{align}
where $G_\Lambda(k)=G_{\Lambda,12}(k)$ and $F_\Lambda(k)=-G_{\Lambda,11}(k)$, and we have neglected terms of second order in $\delta\Gamma^{(3)}_\Lambda$.
This $\delta W^{(2{\rm c})}_\Lambda(k)$ vanishes if we set $(G_{\Lambda},\dot{G}_\Lambda)=(F_{\Lambda},\dot{F}_{\Lambda})$.
Hence, we express $(G_{\Lambda},\dot{G}_\Lambda)=(F_{\Lambda}+\delta G_\Lambda,\dot{F}_{\Lambda}+\delta \dot{G}_\Lambda)$ and retain terms up to the first order in $(\delta G_\Lambda,\delta\dot{G}_\Lambda)$. Finally substituting
\begin{align*}
\delta G(q)=-\frac{\varTheta(q-\Lambda)}{2g_\Lambda \Psi_\Lambda^2},\hspace{5mm} \delta \dot{G}(q)=\frac{\delta(q-\Lambda)}{2g_\Lambda \Psi_\Lambda^2}
\end{align*}
as obtained from Eqs.\ (\ref{hatGdG_Lambda}) and (\ref{calG2}), we arrive at Eq.\ (\ref{W^(2c)-2}).

\end{document}